\documentclass[epjST]{svjour}
\usepackage{longtable}
\usepackage{subfigure}
\usepackage{mathrsfs}
\usepackage{graphics}
\usepackage{epstopdf}
\usepackage{hyperref}
\usepackage{amssymb}
\usepackage{amsmath}
\usepackage{mathrsfs}
\usepackage{array}
\usepackage{graphicx}% Include figure files
\usepackage{dcolumn}% Align table columns on decimal point
\usepackage{bm}% bold math
\usepackage{epstopdf}
\usepackage{tensor}
\usepackage{gensymb} 
\usepackage{cancel}
\usepackage{float}
\usepackage{color}
\usepackage{float}

\begin{document}
\title{Investigation of $\Xi^- nn$  ($S=-2$) Hypernucleus in Low-energy Pionless Halo Effective Theory}
\author{Ghanashyam Meher\inst{1}\fnmsep\thanks{{ghanashyam@iitg.ac.in}} \and Udit Raha\inst{1}\fnmsep\thanks{udit.raha@iitg.ernet.in}}
\institute{\inst{1}Department of Physics, Indian Institute of Technology Guwahati, 781 039 Assam, India}
%
%%%%%%%%%%%%%%%%%%%%%%%%%%%%%%%%%%%%%%%%%%%%%%%%%%%%%%%%%%%%%%%%%%%%%%%%%%%%%%%%%%%%%%%%%%%%%%%%%%%%%%%%%%%%%%%%%%%
%                                                     ABSTRACT
%%%%%%%%%%%%%%%%%%%%%%%%%%%%%%%%%%%%%%%%%%%%%%%%%%%%%%%%%%%%%%%%%%%%%%%%%%%%%%%%%%%%%%%%%%%%%%%%%%%%%%%%%%%%%%%%%%%
\abstract{
We study the $\Xi^- nn$ ($S=-2,\,I=3/2,\,J^P={1/2}^+$) three-body system using low-energy effective field theory (EFT). Due 
to the acute inadequacy of empirical information in this sector, there exists substantial degree of ambiguity in determining 
various few-body observables, some of which are expected to yield vital clues to resolving longstanding contentious issues in 
hypernuclear physics. Moreover, in astrophysical studies, a precise determination of neutron star equation of state (EoS) of 
putative hyperonic cores relies on essential input from the $S=-2$ sector. In this obscure current scenario, a pionless EFT 
analysis provides a systematic model independent framework for assessing the feasibility of light three-particle-stable bound 
states, utilizing low-energy universality. Here we take recourse to a simplistic speculation of the three-body system by 
eliminating the repulsive spin-singlet $\Xi^- n$ sub-system, while retaining the predominantly attractive (possibly bound) 
spin-triplet $\Xi^- n$ and the virtual bound spin-singlet $nn$ sub-systems. In particular, a qualitative leading order EFT 
investigation by introducing a sharp momentum ultraviolet cut-off parameter $\Lambda_{\rm reg}$ into the coupled integral 
equations indicates a discrete scaling behavior akin to a renormalization group limit cycle, thereby suggesting the formal 
existence of Efimov states in the unitary limit, as $\Lambda_{\rm reg}\to \infty$. Our subsequent non-asymptotic analysis 
indicates that the three-body binding energy $B_3$ is sensitively dependent on the cut-off without the inclusion of three-body 
contact interactions. Furthermore, our analysis reproduces several values of the binding energy $B_3\sim 3-4$~MeV, predicted in
context of existing potential models, with the regulator $\Lambda_{\rm reg}$ in the range $\sim 350-460$~MeV. Finally, based 
on these model inputs for $B_3$, a ballpark estimate of the three-body scattering length in the range $2.6-4.9$~fm, is naively 
constrained by our EFT analysis. Despite approximations, the resulting Phillips line is expected to yield a robust feature of the halo-bound 
$\Xi^- nn$ system. For pedagogical reasons, using a simple toy model interacting three-bosons system, we highlight in the 
appendices the typical universal features leading to emergence of RG limit cycle and Efimov states which are amenable to a 
low-energy EFT formalism. 
%Brief remarks in light of the application of $\Xi$-hypernuclei on the neutron star EoS are also included in the appendix for completeness.  
} %end of abstract
%%%%%%%%%%%%%%%%%%%%%%%%%%%%%%%%%%%%%%%%%%%%%%%%%%%%%%%%%%%%%%%%%%%%%%%%%%%%%%%%%%%%%%%%%%%%%%%%%%%%%%%%%%%%%%%%%%%%
\maketitle
\pagebreak

%%%%%%%%%%%%%%%%%%%%%%%%%%%%%%%%%%%%%%%%%%%%%%%%%%%%%%%%%%%%%%%%%%%%%%%%%%%%%%%%%%%%%%%%%%%%%%%%%%%%%%%%%%%%%%%%%%%%
\section{Introduction}\label{intro}
%%%%%%%%%%%%%%%%%%%%%%%%%%%%%%%%%%%%%%%%%%%%%%%%%%%%%%%%%%%%%%%%%%%%%%%%%%%%%%%%%%%%%%%%%%%%%%%%%%%%%%%%%%%%%%%%%%%%
The physics of hypernuclei has gained considerable attention in the strangeness nuclear physics community through numerous 
studies of exotic hypernuclei (for recent reviews, see e.g., 
Refs.~\cite{Petschauer:2020urh,Hammer:2019poc,Hiyama:2018lgs,Gal:2016boi,Gal:2020adf,Garcilazo:2020kiy}). Such studies have 
also proven to have important consequences in the astrophysics of neutron stars where strange matter is expected to appear at 
their cores (e.g., see Refs.~\cite{Tolos:2020aln,Watts:2016uzu,SchaffnerBielich:2000wj}). Especially, the strangeness $S=-2$ 
sector has engendered for a long time a great deal of activity behind ideas, such as the existence of the putative 
{\it H-dibaryon} and other light {\it $\Xi$-hypernuclei} or to seek a resolution to the well-known 
{\it hyperon puzzle}~\cite{Demorest:2010bx,Antoniadis:2013pzd}. For instance, with regard to the feasibility of the 
$H$-particle, as conjectured by Jaffe more than 40 years ago~\cite{Jaffe:1976yi}, no definite conclusion has been reached 
till date, despite the extensive theoretical~\cite{Oka:1983ku,SilvestreBrac:1987cx,Straub:1988mz,Nakamoto:1997gh,Beane:2010hg,Inoue:2010es,Beane:2011iw,Shanahan:2011su,Haidenbauer:2011ah,Inoue:2011ai,Sasaki:2018mzh,Yamaguchi:2016kxa,Francis:2018qch,Garcilazo:2020ofz} 
and experimental investigations~\cite{Ahn:1998fj,Kim:2013vym}. It has been recognized that a thorough understanding of the 
role and character of the underlying $YNN$ and $YYN$ three-body forces (3BFs) is vital towards resolving some of these 
contentious issues. Especially, to stabilize neutron stars with masses larger than twice the solar mass ($2M_{\odot}$) 
against gravitational collapse, the sole inclusion of $NN,\,YN$, and $YY$ two-body interactions becomes questionable, as they 
lead to considerable {\it softening} of the {\it equation-of-state} (EoS)~\cite{Lattimer2010} of the dense baryonic matter. 
The answer probably lies in the inclusion of an admixture of $NNN$, $\Lambda NN$, $\Lambda\Lambda N$ and $\Xi NN$ 3BFs that 
may be the key in estimating the correct stiffness of the EoS governing the stability of the cores. Notably, the 
{\it Quantum Monte Carlo} simulations by Lonardini {\it et al}~\cite{Lonardoni:2013gta,Lonardoni:2014bwa} have already shown
encouraging indication that the inclusion of $\Lambda NN$ 3BF compensates the excessive overbinding due to the $\Lambda N$ 
interactions, ostensibly resolving the ``$B_\Lambda$-overbinding" problem. Further work in this direction is necessary for a 
comprehensive understanding of the 2017 observation of gravitational waves from two-neutron star mergers by the LIGO 
Scientific Collaboration~\cite{TheLIGOScientific:2017qsa}. 

\vspace{0.1cm}

In 2001, the NAGARA event~\cite{Takahashi:2001nm} from the KEK E373 emulsion experiment undoubtedly provided the first 
evidence of the light double-$\Lambda$ hypernuclei ${}^{\,\,\,\,6}_{\Lambda\Lambda}$He, demonstrating that the $S=-2$ 
$\Lambda\Lambda$ interactions are less attractive than the $S=-1$ $\Lambda N$ counterparts. On the other hand, the 
feasibility of a light $\Xi$-hypernuclei based on the state-of-the-art 
experimental~\cite{Khaustov:1999bz,Aoki:2009pvs,Nakazawa:2015joa} and
theoretical~\cite{Hiyama:2008fq,Carames:2012zz,Garcilazo:2015noa,Sun:2016tuf,Garcilazo:2016ylj,Garcilazo:2016ams,Filikhin:2017fog,Hiyama:2019kpw,Jin:2019sqc} 
studies remains largely equivocal since they were first claimed in 1959 in the experimental work of Wilkinson 
{\it et al.}~\cite{Wilkinson:1959zz}. This is primarily due to the acute scarcity of $S=-2$ empirical 
information\footnote{Due to the current impracticability of $\Xi$-hyperon scattering experiments, accurate data is 
difficult to procure. However, study of pertinent correlations in heavy-ion collisions or highly energetic proton-proton 
scattering in future facilities like ALICE and PANDA could certainly improve the present scenario~\cite{Fabbietti:Hyp2018,SanchezLorente:2014jxa}. } needed to determine the underlying character
of the hypernuclear interactions. All that one finds in the literature are a few scattered upper bounds for 
$\Xi^- p\to \Xi^-p$ (elastic) and  $\Xi^- p\to \Lambda \Lambda$ (inelastic) cross sections from emulsion 
experiments~\cite{Aoki:1998sv,Tamagawa:2001tk,Ahn:2005jz}  for laboratory frame momenta in the range $500-600$~MeV. Thus, it 
is no surprise that different existing model analyses lead to substantially contrasting views regarding the nature of the 
$\Xi N$ potentials, ranging from moderately or weakly
attractive~\cite{Khaustov:1999bz,Hiyama:2019kpw,Aoki:1998sv,Tamagawa:2001tk,Ahn:2005jz,Yamaguchi:2001ip,Friedman:2007zza,Hiyama:2010zz,Haidenbauer:2015zqb,Li:2018tbt,Haidenbauer:2018gvg}, 
even vanishing~\cite{Kohno:2009ny}, to weakly repulsive~\cite{Krishichayan:2010zza}. In fact, the KISO 
event~\cite{Nakazawa:2015joa} from the KEK E373 experiment in 2015, which undeniably confirmed the particle-stable
$\Xi$-hypernucleus $^{15}_\Xi$C (interpreted as the ground state of a deeply bound $\Xi^{-}\!\!-\!{}^{14}$N cluster system 
with binding energy $4.38 \pm 0.25$~MeV), at the least corroborated that the constituent $\Xi N$ channels are attractive. 
Specifically, the updated ({\it Extended-soft-core} ESC08c) Nijmegen model {\it G-matrix}
analyses~\cite{Nagels:2015dia,Rijken:2016uon} have predicted that the $\Xi N$ two-body system in the maximal spin-isospin 
$(i=1,j^p=1^+)$ channel is strongly attractive and forms a near-threshold bound state with large positive ${}^3S_1$ 
scattering length, namely, $a^{(j=1)}_{\Xi n}=4.911$ fm. On the contrary, the $(i=1,j^p=0^+)$ channel was predicted to be 
predominantly repulsive with small ${}^1S_0$ scattering length, namely, $a^{(j=0)}_{\Xi n}=0.579$~fm. Using a potential model
involving {\it Faddeev equations}, the putative two-body bound state in the former attractive $\Xi N$ channel, so-called the 
{\it deuteron*} ($D^*$), was estimated to have a binding energy of $1.56$~MeV ($1.67$~MeV) with (without) taking into 
consideration the latter repulsive $\Xi N$ channel~\cite{Garcilazo:2016ams,Filikhin:2017fog}. In a contrasting scenario, 
the recent SU(3) chiral effective field theory (EFT) predictions from the relativistic leading order (LO) analysis of 
Ref.~\cite{Li:2018tbt}, as well as the non-relativistic {\it next-to-leading order} (NLO) {\it in-medium} G-matrix analysis 
of Ref.~\cite{Haidenbauer:2018gvg}, have practically ruled out the possibility of a particle-stable $\Xi N$ bound state in 
the (1,1) channel. It is noteworthy that both EFT analyses were constrained by the recent HAL QCD lattice 
results~\cite{Sasaki:2019qnh} and the aforementioned empirical upper bounds from $\Xi^- p$ cross sections
data~\cite{Aoki:1998sv,Tamagawa:2001tk,Ahn:2005jz,Haidenbauer:2015zqb}. Interestingly, the recent Faddeev 
calculations~\cite{Garcilazo:2020kiy,Garcilazo:2020ofz,Garcilazo:2015noa,Garcilazo:2016ylj,Garcilazo:2016ams,Filikhin:2017fog} 
relying either on the updated $\Xi N$ Nijmegen ESC08c potential model~\cite{Nagels:2015dia,Rijken:2016uon} as input for the 
$I=3/2,\,J^P={1/2}^+$ channel, or on the recent HAL QCD based $\Lambda\Lambda-\Xi N$ separable interaction 
potential~\cite{Sasaki:2019qnh} as input for the $I=1/2,\,J^P={1/2}^+$ channel, have hinted at the feasibility of a deeply 
bound $\Xi NN$ state in the former channel and a three-body $\Xi NN-\Lambda \Lambda N$ resonance state (i.e., either as a 
$\Lambda\Lambda N$ resonance state or a $\Xi NN$ quasi-bound state) in the latter.\footnote{Even a contrasting viewpoint is 
obtained in the $\Xi NN$ system in the light of the more recent variational calculation using {\it Gaussian expansion} 
method~\cite{Hiyama:2019kpw} with Nijmegen $\Xi N$ potential~\cite{Nagels:2015dia,Rijken:2016uon}. Oddly enough, their findings
indicate a rather strongly attractive $I=1/2,\,J^P={1/2}^+$ channel with a three-body bound state with binding energy $7.20$~MeV,
while the $I=3/2,\,J^P={1/2}^+$ channel is less attractive without a bound state. }  These facts ostensibly imply that the $\Xi N$ 
interactions are predominantly attractive in nature.

%%%%%%%%%%%%%%%%%%%%%%%%%%%%%%%%%%%%%%%%%%%%%%%%%%%%%%%%%%%%%%%%%%%%%%%%%%%%%%%%%%%%%%%%%%%%%%%%%%%%%%%%%%%%%%%%%%%%
\section{Pionless EFT (${}^{{\pi}\!\!\!/}$EFT): A Brief Survey}
%%%%%%%%%%%%%%%%%%%%%%%%%%%%%%%%%%%%%%%%%%%%%%%%%%%%%%%%%%%%%%%%%%%%%%%%%%%%%%%%%%%%%%%%%%%%%%%%%%%%%%%%%%%%%%%%%%%%
Prompted by this unresolved scenario, we present in this work an alternative qualitative assessment regarding 
the viability of a putative $\Xi^{-} nn$ two-neutron {\it halo-bound} state in the $I=3/2,\, J^P={1/2}^+$ channel. 
In particular, the reasons motivating our study of the $\Xi NN$ system in this maximal spin-isospin channel are 
as follows:
\begin{itemize}
\item First, the decoupling of this channel from the strong decay into the $\Lambda\Lambda N$ channel is 
forbidden by isospin conservation.\footnote{The $\Xi NN$ system in the $I=1/2,\, J^P={1/2}^+$ channel, on the 
other hand, has profound astrophysical importance in the context of EoS of neutron star matter. It has been 
recognized that $\Xi N N\to \Lambda\Lambda N$ transmutations could contribute to an intricate balance between 
the ordinary nucleonic and hyperonic matter accumulating at the stellar cores, inducing a natural ``pressure 
control" mechanism for the build-up of neutron and lepton Pauli pressures in high-density matter. Moreover, the
$\Lambda NN$ and $\Lambda\Lambda N$ three-body observables can additionally serve to fine-tune the stiffness of 
the EoS in a controlled way. In this regard, a series of chiral constituent quark model analyses by Garcilazo 
{\it el al.}~\cite{Garcilazo:2012qv,Garcilazo:2014pxa,Garcilazo:2014mra} using Faddeev equations suggested the 
importance of $\Xi NN -\Lambda\Lambda N$ couplings in obtaining a three-body bound state, so-called the 
$S=-2$ {\it hypertriton} (with binding energy $\sim 0.5$~MeV), given that the $\Lambda\Lambda N$ system is by 
and large unbound. However, such a bound state mechanism seems fundamentally at odds with Efimov universality, 
since the feasibility of Efimov states gets substantially weakened or disappears in proximity to open decay or 
reaction channels~\cite{Hyodo:2013zxa,Raha:2017ahu}. Thus, it seems rather unlikely that the above reported 
bound state is manifestly Efimov-like in character. This calls for a rigorous model independent assessment which
is beyond the scope of a simplistic qualitative treatment as pursued in this work. } \\
\item Second, Pauli principle works favorably in supporting stable $\Xi NN$ bound states. \\
\item Third, the absence of Coulomb effects to a great extent simplifies the EFT construction of the coupled 
integral equations in the momentum-space~\cite{Braaten:2004rn}, the so-called STM or  
{\it Skornyakov-Ter-Martirosyan} equations~\cite{STM1,STM2} (cf. Appendix A.2). 
\end{itemize}
Our treatment is based on a low-energy {\it pionless} EFT
(${}^{{\pi}\!\!\!/}$EFT)~\cite{Braaten:2004rn,Bedaque:1998kg,Bedaque:1999ve,Bedaque:1998km,Kaplan:1996nv,Kaplan:1996xu,Kaplan:1998tg,Kaplan:1998we,Kaplan:1998sz,vanKolck:1998bw,Bedaque:1998mb,Birse:1998dk,Beane:2000fx} where explicit pion exchanges are integrated out at
scales much smaller than the pion mass. A speciality of such an approach is that the results are obtained following a general 
model-independent perturbative scheme utilizing principles of low-energy universality with {\it controlled} error estimates. 
Observables are expressed as an expansion of a small low-energy parameter $\epsilon=Q/\Lambda_H$,with $Q$ being the typical 
momentum scale of dynamics of the system in question, and $\Lambda_H\sim m_\pi$ is the {\it hard} or {\it breakdown} scale 
of the theory which is identified with the pion mass $m_\pi$. Such a methodology is complementary to {\it ab initio} approaches,
where the universal phenomenological couplings or {\it low-energy constants} (LECs) in the effective Lagrangian could be used to 
make predictions on various few-body observables. Such universal aspects of ${}^{{\pi}\!\!\!/}$EFT have been successfully 
exploited to investigate the dynamics of {\it finely tuned} systems of atoms and light nuclei driven arbitrarily close to the 
unitary limit of two-body short-distance interactions. This is either achieved {\it artificially}, by tuning inter-atomic 
potentials in selective open channels using varying electromagnetic fields, as in {\it Feshbach resonances}~\cite{Braaten:2004rn} 
in ultra-cold atoms, or even {\it naturally}, as in nuclear systems with large two-body scattering lengths. This leads to 
formation of threshold two-body bound states, such as in the case of the {\it deuteron} ($np$ bound state) or in our context of the aforementioned 
putative bound $D^*$ state~\cite{Nagels:2015dia,Rijken:2016uon}. More interestingly, interacting three-body S-wave systems when 
driven in proximity to the unitary limit, lead to the well-known {\it Efimov
phenomenon}~\cite{Braaten:2004rn,Danilov:1961,Efimov:1970zz,Efimov:1971zz,Efimov:1973awb,Naidon:2016dpf}, associated with an 
infinite tower of arbitrarily shallow geometrically spaced three-body levels accumulating to zero-energy, namely, the
{\it three-particle} or {\it particle-dimer} break-up threshold (see, e.g., \cite{Hammer:2019poc,Braaten:2004rn,Naidon:2016dpf} 
and reference therein for a detailed review of Efimov physics and its applications in atomic and nuclear physics). In that case a 
modified ${}^{{\pi}\!\!\!/}$EFT {\it power counting} scheme was suggested by Bedaque {\it et al.}~\cite{Bedaque:1998kg,Bedaque:1999ve}. 
The power counting mandates {\it non-derivative} three-body contact interaction couplings, which are otherwise subleading in a 
{\it naive} dimensional analysis (NDA), to be promoted to the LO whenever they exhibit renormalization group (RG) {\it limit 
cycle}. For pedagogical purpose, Appendix A highlights brief technical details of the ${}^{{\pi}\!\!\!/}$EFT formalism capturing 
the two- and three-body universal physics relating to Efimov-like bound states. A simple toy model analysis of a system of three 
identical interacting bosons provides the essential background for the methodology adopted in this paper. 

\vspace{0.1cm}

An important variant of the standard ${}^{{\pi}\!\!\!/}$EFT-based on the generalization of nuclear cluster models is the so-called 
{\it halo/cluster} ${}^{{\pi}\!\!\!/}$EFT~\cite{Bertulani:2002sz,Bedaque:2003wa}. It was primarily introduced for investigating the
{\it clustering} and {\it halo} phenomena in light nuclei with narrow resonances (often in higher partial waves), characterized by 
multi-scale threshold dynamics at energy scales often lower than that in standard ${}^{{\pi}\!\!\!/}$EFT. A {\it heteronuclear} 
subclass of systems often manifest themselves as exotic s-shell {\it hypernuclei} which typically lie along the limits of nuclear 
stability (so-called the {\it driplines}). These modified ${}^{{\pi}\!\!\!/}$EFTs exploit the separation of scales between  the hard 
scale of the EFT and a hierarchy of dynamically generated low-energy scales associated with the formation of one or more {\it shallow} 
bound/resonance states. Such analyses has been successfully applied to study light hypernuclei, since the very first of such 
EFT work by Hammer~\cite{Hammer:2001ng} on the LO investigation of hypertriton (${}^3_\Lambda$H), a $\Lambda np$ Efimov-like bound 
system in the $I=0, J=1/2$ channel. Subsequently, a number of similar LO halo/cluster EFT works appeared in the literature, both in 
the $S=-1$~\cite{Ando:2015fsa,Hildenbrand:2019sgp} and $S=-2$~\cite{Ando:2013kba,Ando:2014mqa,Meher:2020rsd} strangeness sectors, in 
the search for light exotic single and double $\Lambda$-hypernuclear states, e.g., $nn\Lambda$,\footnote{The ${}^{{\pi}\!\!\!/}$EFT 
analysis by Ando {\it et al.}~\cite{Ando:2015fsa} attempted to investigate the feasibility of the putative $nn\Lambda$ bound state, 
as reported by the HypHI Collaboration~\cite{Rappold:2013jta} in 2013. In that analysis a coupled system of integral equation was 
constructed in the physical basis involving only the spin projected couplings and excluding isospin projections for simplicity. This, however, yielded the 
asymptotic RG limit cycle scaling exponent as $s^{(nn\Lambda)}_0=0.80339...$, which is inconsistent with the expected universal 
scaling based on the the relative three-particle mass ratios~\cite{Braaten:2004rn}. As recently elucidated by Hildenbrand and
Hammer~\cite{Hildenbrand:2019sgp}, the correct scaling could be achieved by a proper reformulation in the spin-isospin basis 
leading to the value, $s^{(nn\Lambda)}_0=1.0076\cdots$. This value is identical to that obtained in the study of hypertriton, and also 
reproduces the well-known asymptotic scaling $s_0=1.00624\cdots$, for identical masses~\cite{Braaten:2004rn,Naidon:2016dpf} (also 
see Appendix A.3). Furthermore, in Ref.~\cite{Hildenbrand:2019sgp} a threshold ground state appeared at the critical cut-off scale 
$\Lambda_{\rm reg}\sim 600$~MeV, whereby the likelihood of a physically realizable Efimov-bound/resonance $\Lambda nn$ state may 
not be excluded outright. Notably, such a possibility had been completely ruled out earlier in Ref.~\cite{Ando:2015fsa} with the 
critical cut-off obtained as $\Lambda_{\rm reg} \gtrsim 1.5$~GeV. Besides, it deserves mentioning here that nearly all potential 
model approaches till date have reported negative results for the existence of the $\Lambda nn$ bound state (see e.g., 
Refs.~\cite{Gal:2014efa,Garcilazo:2014lva,Hiyama:2014cua,Afnan:2015ahc}). In particular, the Faddeev calculation analysis of
Ref.~\cite{Afnan:2015ahc} demonstrated using a {\it complex scaling} method that the strength of the $\Lambda n$ Yamaguchi-type 
(separable) potential is needed to be tuned $\sim 25\%$ above the realistic estimate in order for the $\Lambda nn$ system to 
emerge into a three-body bound state. } 
${}^{\,\,\,\,\,4}_{\Lambda\Lambda}{\rm He}$,\, ${}^{\,\,\,\,\,5}_{\Lambda\Lambda}{\rm H}$,\,
${}^{\,\,\,\,\,5}_{\Lambda\Lambda}{\rm He}$ and ${}^{\,\,\,\,\,6}_{\Lambda\Lambda}{\rm He}$. It is worth mentioning here that a 
novel {\it ab initio} LO ${}^{{\pi}\!\!\!/}$EFT technique using few-body {\it stochastic variational} method of calculation was 
suggested in Refs.~\cite{Barnea:2013uqa,Kirscher:2015yda,Contessi:2017rww,Kirscher:2017fqc} for the study of ordinary nuclei on 
the lattice, the so-called {\it lattice nuclei}, facilitating easy comparison with results of Lattice QCD simulations at 
unphysical quark masses. Such a framework, which is complimentary to the halo ${}^{{\pi}\!\!\!/}$EFT approach, was later extended
by Contessi {\it et al.}~\cite{Contessi:2018qnz} in the $S=-1$ sector to seek a solution to the $B_\Lambda$-overbinding problem, 
and more recently in the feasibility studies of several light $S=-2$ double $\Lambda$-hypernuclei~\cite{Contessi:2019csf}. Here 
we re-emphasize that the above literature survey relating to studies on $\Lambda$-hypernuclei comprises only of noteworthy 
${}^{{\pi}\!\!\!/}$EFT works motivated on the philosophy of few-body universality, as adopted in this paper. Needless to say, 
however, that the existing literature also includes an entire gamut of well acclaimed works based on {\it ab initio} and 
{\it cluster} potential models involving three- and four-body Faddeev-Yakubovsky and variational calculations (see e.g., 
\cite{Hiyama:2018lgs,Gal:2016boi,Gal:2020adf,Garcilazo:2020kiy} and references therein). As these methodologies do not come under 
the direct purview of universal physics principles,  a detailed discussion of the models, especially in the context of 
$\Lambda$-hypernuclear studies, is beyond the scope of this work.   

%%%%%%%%%%%%%%%%%%%%%%%%%%%%%%%%%%%%%%%%%%%%%%%%%%%%%%%%%%%%%%%%%%%%%%%%%%%%%%%%%%%%%%%%%%%%%%%%%%%%%%%%%%%%%%%%%%%%
\section{Halo ${}^{{\pi}\!\!\!/}$EFT of $\Xi^- nn$}\label{sec:1}
%%%%%%%%%%%%%%%%%%%%%%%%%%%%%%%%%%%%%%%%%%%%%%%%%%%%%%%%%%%%%%%%%%%%%%%%%%%%%%%%%%%%%%%%%%%%%%%%%%%%%%%%%%%%%%%%%%%%
In this work, we use halo ${}^{{\pi}\!\!\!/}$EFT at LO to assess the feasibility of a $\Xi^- nn$  bound state in the $I=3/2,\,J=1/2$ 
channel primarily based of Efimov universality. This may be reflected through a study of the EFT regulator scale dependence of the 
RG limit cycle exhibited by the three-body contact interaction coupling (see Appendix A for basic details regarding few-body 
universality, RG limit cycle and Efimov effect in the context of EFT analysis). Despite existing uncertainties regarding the exact 
nature of $\Xi N$ interactions, we adopt a certain scenario motivated by the results from a series of recent 
{\it constituent quark cluster} potential model (CQCM)
analyses~\cite{Garcilazo:2015noa,Garcilazo:2016ylj,Garcilazo:2016ams,Filikhin:2017fog} based on Faddeev calculations. These analyses 
rely on the fact that the ${}^{3}S_1$ $\Xi^{-} n$ sub-system is dominantly attractive and bound with large {\it positive} S-wave 
scattering length, namely, $a^{(j=1)}_{\Xi n}=4.911$~fm, taken from the Nijmegen ESC08c model~\cite{Nagels:2015dia,Rijken:2016uon}. 
It is especially noted in some of these model studies that if one included the real bound ${}^{3}S_1$ channel as the only $\Xi^- n$ 
sub-system channel, the $\Xi^-nn$ system exhibited a three-body deeply bound state. On the other hand, no bound state is obtained 
with only the ${}^1S_0$ $\Xi^- n$ sub-system channel included with the Nijmegen model predicted small S-wave scattering length, namely,
$a^{(j=0)}_{\Xi n}=0.579$~fm~\cite{Nagels:2015dia,Rijken:2016uon}.\footnote{It is notable that the ${}^{1}S_0$ $nn$ sub-system 
channel is {\it virtual} bound with large {\it negative} S-wave scattering length, namely, $a_{nn}=-18.63$~fm~\cite{Chen:2008zzj}.}. 
Here we report analogous qualitative features arising in the context of our EFT framework as well. This probably hints at the 
consistency of our halo EFT results with those reported earlier in the above-mentioned  model analyses. In particular, the Faddeev-type coupled
integral equations in the momentum space are found not to exhibit an RG limit cycle with only the repulsive ${}^1S_0$ $\Xi^- n$ 
sub-system channel included, implying an unbound $\Xi^- nn$ system. On the other hand, we find that an {\it asymptotic} RG limit 
cycle (cf. Appendix A.3) is always manifest in the presence of the $\Xi^- n$ triplet channel, irrespective of the 
inclusion of the $\Xi^- n$ singlet channel. However, a full-fledged numerical evaluation of the integral equations using conventional
auxiliary fields, or the so-called {\it dibaryon} formalism~\cite{Bedaque:1998kg,Bedaque:1999ve,Birse:1998dk} (cf. Appendix A.1 for 
details), becomes a challenging task, especially when dealing with the repulsive $\Xi^- n$ sub-system with a small positive scattering
length. The problem is attributed to the presence of a unphysically deep pole in the ${}^1S_0$ $\Xi^- n$ dibaryon propagator [cf.
Eq.~\eqref{eq:dimer_props}] corresponding to an unnaturally large binding momentum, 
$\gamma^{(0)}_{\Xi n}\approx 1/a^{(0)}_{\Xi n}\approx 340$~MeV (estimated using the the Nijmegen ESC08c
model~\cite{Nagels:2015dia,Rijken:2016uon}). Since there is no straightforward way of ``renormalizing" the effect of such a deep 
two-body pole, the EFT evidently breaks down. Thus, in this work we take recourse to a simplistic study of the $\Xi^- nn$ system to look for possible emergence of physically realizable Efimov-like trimers by completely
excluding the repulsive ${}^{1}S_0$ $\Xi^- n$ channel (as done in e.g., Ref.~\cite{Garcilazo:2015noa}). Notably, in our halo EFT 
formalism the triplet dibaryon pole position defines the $n+(\Xi^- n)_t$ particle-triplet-dimer break-up threshold, beyond which the
$\Xi^- nn$ trimer levels are expected to emerge. The trimer binding energies correspond to the eigensolutions to the coupled integral 
equations for a given finite value of an ultraviolet (UV) sharp momentum cut-off regulator~\cite{Braaten:2004rn} (cf. Appendix A.2).

\vspace{0.1cm}

In the ensuing EFT analysis, we present a qualitative investigation of the regulator scale dependence of an {\it a priori} 
undetermined three-body contact interaction coupling introduced for the purpose of renormalization. Through this study we 
hope to establish a correspondence between our EFT results with those obtained in the potential model analyses by Garcilazo 
{\it et al.}~\cite{Garcilazo:2015noa,Garcilazo:2016ylj,Garcilazo:2016ams,Filikhin:2017fog}. Our analysis serves as a consistency 
check between both types of approaches. In particular, based on existing model estimates for the three-body binding energy, 
we give a {\it naive} window of possible estimates of the S-wave three-body scattering length associated with the elastic 
$n-(\Xi^-n)_t$ scattering process. These predictions in turn induce a striking feature of three-body universality, the 
so-called {\it Phillips line}~\cite{Phillips:1968zze}, namely, the fact that different model potentials tuned to the same 
input two-body scattering data (i.e., $a^{(j=1)}_{\Xi n}$ and $a_{nn}$) yield a highly correlation result for the $\Xi^- nn$ 
binding energy and the corresponding three-body scattering length. Albeit approximations considered in our simplistic 
model independent treatment, such universal correlations are expected to reflect robust features of the thee-body system.

%%%%%%%%%%%%%%%%%%%%%%%%%%%%%%%%%%%%%%%%%%%%%%%%%%%%%
\subsection{Effective Lagrangian and Formalism}
%%%%%%%%%%%%%%%%%%%%%%%%%%%%%%%%%%%%%%%%%%%%%%%%%%%%%
In a simplified picture, the $\Xi^- nn$ system may be visualized as a two-neutron halo with the two loosely bound neutrons 
orbiting about the $\Xi^-$-hyperon ``elementary" core, forming a shallow bound state with a diffuse structure. Such universal 
class of systems exploits the distinct separation of scale between the typical dynamical scale, 
$Q\sim\gamma^{(1)}_{\Xi n}\approx 1/a^{(1)}_{\Xi n}\sim 40$~MeV, associated with the ``attractive pole" momentum of the 
${}^3S_1$ $\Xi^- n$ dibaryon propagator (ignoring possible artifact due to the deep pole at 
$\gamma^{(0)}_{\Xi n}\approx 1/a^{(0)}_{\Xi n}\sim 340$~MeV associated with the repulsive ${}^1S_0$ $\Xi^- n$ sub-system), 
and the breakdown scale $\Lambda_H\sim m_\pi$ of standard ${}^{{\pi}\!\!\!/}$EFT~\cite{Braaten:2004rn,Bedaque:1998kg,Bedaque:1999ve,Bedaque:1998km,Kaplan:1996nv,Kaplan:1996xu,Kaplan:1998tg,Kaplan:1998we,Kaplan:1998sz,vanKolck:1998bw,Bedaque:1998mb,Birse:1998dk,Beane:2000fx}. This implies that $\epsilon \sim Q/m_\pi \sim 1/3$ 
defines a reasonable expansion parameter that is amenable to a EFT treatment. A concise description on the general principles 
and methodology of ${}^{{\pi}\!\!\!/}$EFT framework used in the analyses of two- and three-body universality is provided in 
Appendix A. The effective Lagrangian is constructed on the basis of all possible available low-energy symmetries 
(${\cal P},\, {\cal C},\, {\cal T}$ and Galilean invariance) and degrees of freedom. The interaction vertices are represented 
by local contact interactions and the Lagrangian is expressed in a derivative expansion of the fundamental fields. For our 
system, the fundamental degrees of freedom consist of the $\Xi^-$-hyperon and neutron ($n$) fields. The LO non-relativistic 
Lagrangian is free from derivative terms and expressed as a sum of one-, two- and three-body parts, namely, 
\begin{equation}
\mathcal{L}_{{}^{{\pi}\!\!\!/}{\rm EFT}} = \mathcal{L}_{\rm 1-body} + \mathcal{L}_{\rm 2-body} + \mathcal{L}_{\rm 3-body}\,\,.
\label{eq:1}
\end{equation}
Below we consider each of the components of the effective Lagrangian separately. \\

\noindent{\it One-body part.\,\,}
The terms $\mathcal{L}_{\Xi}$ and $\mathcal{L}_n$ constitute the one-body Lagrangian $\mathcal{L}_{\rm 1-body}$ corresponding 
to the kinetic part of the $\Xi^-$-hyperon and neutron fields respectively, and are expressed in the physical basis as 
\begin{eqnarray}   
\mathcal{L}_{\rm 1-body} = \mathcal{L}_{\Xi} + \mathcal{L}_{n}\,\,, 
\end{eqnarray}
where
\begin{eqnarray}
\mathcal{L}_{\Xi} &=& {\Xi}^{\dagger}\bigg[iv\cdot\partial+\frac{(v\cdot\partial)^2-\partial^2}{2M_{\Xi}}\bigg]{\Xi}\,\,,
\nonumber \\ 
\mathcal{L}_{n}  &=&  n^{\dagger}\bigg[iv\cdot\partial+\frac{(v\cdot\partial)^2-\partial^2}{2M_n}\bigg]n\,\,,
\end{eqnarray}
where $M_{\Xi,}$ and $M_{\,n}$ are the physical masses of the $\Xi^-$-hyperon and neutron fields respectively, as given in Table~\ref{tab:1}, 
and $v^{\mu}=(1,{\bf 0})$ is the four-velocity vector which is used to express the Lagrangian in a manifestly covariant 
manner akin to the heavy-baryon formalism~\cite{Bernard:1995dp}. It follows that the non-relativistic propagators 
associated with these fundamental fields are given by
\begin{eqnarray}
 iS_{\Xi}(p_0, {\bf p}) &=& \frac{i}{p_0-\frac{\bf p^2}{2 M_{\Xi}}+i\eta}\,\,,
 \nonumber \\
 iS_n (p_0, {\bf p}) &=& \frac{i}{p_0-\frac{\bf p^2}{2 M_n}+i\eta}\,; \quad \eta\to 0\,\,,
\end{eqnarray}
where $p_0$ and ${\bf p}$ are temporal and spatial parts of the generic four-momentum $p^\mu$. \\
%%%%%%%%%%%%%%%%%%%%%%%%%%%%%%%%%%%%%%%%%%%%%%%%%%%%%%%%%%%%%%%%%%%
\begin{table}[tbp]
\begin{center}
\caption{PDG~\cite{Zyla:2020zbs} values of particle masses considered in the analysis. }
\label{tab:1}    
\begin{tabular}{|c|c|c|}
\hline
Particle & Mass Symbol  & Numerical Value (MeV)\\
\hline\hline
$\Xi$-Hyperon & $M_{\Xi}$ & 1321.710  \\
Neutron (n) & $M_n$ & 939.565 \\
\hline\hline
\end{tabular}
\end{center}
\end{table}
%%%%%%%%%%%%%%%%%%%%%%%%%%%%%%%%%%%%%%%%%%%%%%%%%%%%%%%%%%%%%%%%%%%%

\begin{figure}[tbp]
\centering
\includegraphics[width=11cm]{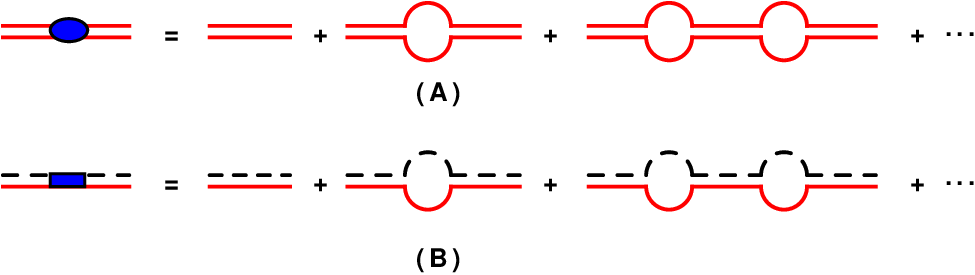}
\caption{The renormalized  dressed  propagators for (upper pannel) ${}^1S_0$ $nn$, and  (lower pannel) ${}^3S_1$ $\Xi^- n$ dibaryon fields. 
         The dashed lines represent the $\Xi^-$-hyperon field propagator and the solid lines represent the neutron field 
         propagator.}
\label{fig-7}
\end{figure}

\noindent{\it Two-body part.\,\,}
In ${}^{{\pi}\!\!\!/}$EFT, to deal with the formation of shallow S-wave bound states one needs to {\it unitarize} the 
two-body sector by employing the so-called Kaplan-Savage-Wise (KSW) power counting
rule~\cite{Kaplan:1996xu,Kaplan:1998tg,Kaplan:1998we,Kaplan:1998sz,vanKolck:1998bw}.
To efficiently capture such two-body physics in the vicinity of a  {\it non-trivial fixed-point} described by the RG of 
the two-body contact interactions, it was suggested to introduce auxiliary {\it dimer} fields in the effective
Lagrangian~\cite{Braaten:2004rn,Bedaque:1998kg,Bedaque:1999ve,Bedaque:1998mb,Birse:1998dk} (also, see Appendix A.1). Thus, 
for the heteronuclear $\Xi^{-}nn$ system we need to introduce two types of dimer fields, namely, the isospin-spin triplet 
($i=1,j=1$) $\Xi^- n$ dibaryon field $u_1$ and the isospin-triplet spin-singlet ($i=1,j=0$) $nn$ dibaryon field $u_0$. 
Here we re-emphasize that the  iso-triplet spin-singlet ($i=1,j=0$) $\Xi^- n$ sub-system channel is considered 
decoupled from the picture as its physics lies beyond the realm of our halo EFT formalism. The corresponding two-body LO Lagrangian
(written in the physical basis) in terms of the dibaryon fields is expressed as
\begin{eqnarray}
\mathcal{L}_{\rm 2-body} = \mathcal{L}_{u_0} + \mathcal{L}_{u_1}\,\,, 
\end{eqnarray}
where
\begin{eqnarray}
\mathcal{L}_{u_0} &=& -(u_0)^{a\dagger}\bigg[iv\cdot\partial+\frac{(v\cdot\partial)^2-\partial^2}{4M_n}\bigg](u_0)^a 
-y_0\!\bigg[(u_0)^{a\dagger}\left(n^{T}\hat{\mathscr P}_{(nn)}^{(1,0)\,a}\,n\right)+ {\rm h.c.}\bigg]\,\,,
\nonumber\\
\nonumber\\
\mathcal{L}_{u_1} \!&=&\! -({\bf u}_1)_k^{a\dagger}\bigg[iv\cdot\partial+\frac{(v\cdot\partial)^2
-\partial^2}{2(M_{\Xi}+M_n)}\bigg]\!({\bf u}_1)^a_k-y_1\!\bigg[({\bf u}_1)_k^{a\dagger}
\left(n^{T}\hat{\mathscr P}_{({\Xi} n)\,k}^{(1,1)\,a}\,{\Xi}\right)+ \rm{h.c.}\bigg]\,\,,
\nonumber\\
\end{eqnarray}
noting that the ``wrong signs" in front of the respective kinetic terms suggest the non-dynamical or quasi-particle nature 
of the dibaryon fields. Here, $\hat{\mathscr P}_{({\Xi} n)\,k}^{(1,1)\,a} = \frac{1}{2} \tau^2\tau^a\sigma_2\sigma_k$, and 
$\hat{\mathscr P}^{(1,0)\,a}_{(nn)}= \frac{1}{\sqrt{8}}\tau^2\tau^a\sigma_2$ are the spin-isospin projection operators, 
with $\sigma_k$ and $\tau^a\,(k,a=1,2,3)$ being the Pauli matrices in the spin and isospin spaces respectively. The 
two-body non-derivatively coupled LO contact interactions or LECs $y_{0,1}$ between the respective dibaryons and their 
constituent elementary fields encode all UV physics that remain unresolved in the EFT. These couplings are easily fixed by
the prescription given in Ref.~\cite{Griesshammer:2004pe}: 
\begin{equation}
 y_1=\sqrt{\frac{2\pi}{\mu}}\,, \quad{\rm and}\quad y_0=\sqrt{\frac{4\pi}{M_n}}\,,
\label{eq:y01}
\end{equation}
where $\mu=M_n M_\Xi/(M_n + M_\Xi)=549.174$~MeV is the reduced mass of $\Xi^- n$ two-body sub-system. Next, we spell out
the renormalized ``dressed" (unitarized) propagators for the $nn$ and $\Xi^- n$ dibaryon fields (cf. Fig.~\ref{fig-7}) 
consistent with the KSW power counting scheme~\cite{Kaplan:1996xu,Kaplan:1998tg,Kaplan:1998we,Kaplan:1998sz,vanKolck:1998bw}:
\begin{eqnarray}
 i{\mathscr D}_0(p_0,{\bf p}) &=& 
\frac{4\pi}{y_0^2 M_n}\frac{i}{\gamma^{(0)}_{nn}-\sqrt{-M_n(p_0-\frac{ {\bf p}^2}{4M_n})-i\eta}-i\eta}\,\,, 
\nonumber \\
 i{\mathscr D}_1(p_0,{\bf p}) &=& 
\frac{2\pi}{y_1^2\mu}\frac{i}{\gamma^{(1)}_{\Xi n}
-\sqrt{-2\mu (p_0-\frac{ {\bf p}^2}{2(M_n+M_{\Xi})})-i\eta}-i\eta}\,; \quad \eta\to 0\,\,,
\label{eq:dimer_props}
\end{eqnarray}
where at the LO in ${}^{{\pi}\!\!\!/}$EFT, we have  $\gamma^{(0)}_{nn} \to 1/a_{nn}$ and $\gamma^{(1)}_{\Xi n} \to 1/a^{(1)}_{\Xi n}$,
as the (1,0) $nn$ and (1,1) $\Xi^-n$  dibaryon (virtual or real bound) binding momenta respectively. It may be noted that the two 
scattering lengths are the only two-body input parameters in our LO EFT. Other parameters, such as S-wave effective range $r_{nn}$ and
$r_{\Xi n}$ formally contribute at NLO in the KSW power counting scheme, which is beyond the scope of this work. \\
%%%%%%%%%%%%%%%%%%%%%%%%%%%%%%%%%%%%%%%%%%%%%%%%%%%%%%%%%%%%%%%%%%%%%%%%%%%%%%%%%%%%%%%%%%%%
%%%%%%%%%%%%%%%%%%%%%%%%%%%%%%%%%%%%%%%%%%%%%%%%%%%%%%%%%%%%%%%%%%%%%%%%%%%%%%%%%%%%%%%%%%%%

\noindent{\it Three-body part.\,\,}
Formally the $\Xi^- nn$ three-body system with a possible fine-tuned two-body sector exhibits the well-known Efimov effect 
close to the unitary or resonant limit of the two-body interactions. This is reflected by the fact that the integral 
equations for the system with only two-body contact interactions become ill-defined in the asymptotic UV limit. The inherent
reason for this anomalous UV behavior is the partial breakdown of the expected {\it fixed-point scaling} invariance of the 
system in the vicinity of bound states into the onset of a {\it discrete scaling} symmetry. A possible remedy to this problem,
for instance, may be obtained by including a sharp momentum UV cut-off regulator $\Lambda_{\rm reg}$ in the integral equations,
thereby, formally introducing another free parameter in the LO EFT (in addition to the two S-wave scattering lengths in the 
two-body sector). This simultaneously necessitates the introduction of LO three-body contact interactions (3BFs) as counterterms with 
scale dependent couplings, such as $g_3(\Lambda_{\rm reg})$, to renormalize the artificial cut-off dependence. The resulting 
atypical scaling behavior gets reflected through the emergence of a RG limit cycle behavior in the 3BF couplings (for a 
pedagogical review on this topic, see Ref.~\cite{Braaten:2004rn}; also see Appendices A.2 and A.3). Here we present a certain 
choice of the LO three-body Lagrangian consistent with the reparametrization symmetries of the coupled system of integral 
equation, and given by
\begin{eqnarray}
 \mathcal{L}_{\rm 3-body} &=&-\frac{g_3(\Lambda_{\rm reg})}{\Lambda_{\rm reg}^2}\Bigg{[}\frac{M_{\Xi} y_1^2}{2} 
 \Big{\{}{(\bf u}_1)_l^a \,\,\hat{\mathcal{P}}_l^{ab} \,\,n\Big{\}}^{\dagger}
 \Big{\{}({\bf u}_1)_k^c \,\,\hat{\mathcal{P}}_k^{cb}\,\,n\Big{\}}
 \nonumber\\
 &&\hspace{2cm}-\frac{\sqrt{3}M_{n} y_0y_1}{\sqrt{2}} \Big{\{}({\bf u}_1)_l^a\,\,
 \hat{\mathcal{P}}_l^{ab} \,\,n\Big{\}}^{\dagger}
 \Big{\{}u_0^c \,\,\hat{\mathcal{P}}^{cb} \,\,{\Xi}\Big{\}}+ {\rm h.c.}\Big{]},\,\quad\,
\label{eq:type_3body}
\end{eqnarray}
where the spin-isospin projection operators have the following forms:
\begin{eqnarray}
  \left[\hat{\mathcal{P}}_k^{cb}\right]_{\alpha\beta} &=& 
  \frac{1}{3\sqrt{3}}\Big{[}(\tau^c\tau^b)_{\alpha\beta}+\delta_{cb}\delta_{\alpha\beta}\Big{]}\sigma_k\,\,,
  \nonumber\\
\left[\hat{\mathcal{P}}^{cb}\right]_{\alpha\beta} &=& \frac{1}{3}\Big{[}(\tau^c\tau^b)_{\alpha\beta}
+\delta_{cb}\delta_{\alpha\beta}\Big{]}\,\,,
\end{eqnarray}
with $\alpha,\beta=1,2$ being the isospin-1/2 SU(2) indices~\cite{Wilbring:2016bda}. The cut-off dependence of 
$g_3$ is {\it a priori} undetermined in the EFT and can be fixed only using a three-body datum, e.g., the three-body
binding energy $B_3$ or the corresponding scattering length $a_3$.\footnote{The three-body datum in this case is 
analogous to the information on parameters, such as $\Lambda_*$ or $\kappa_*$, in addition to the two-body scattering 
length $a_{0}$, necessary for the description of the Efimov spectrum, as detailed in Appendix A.3 for a three-boson
system.} However, none of these empirical information is available currently, either from experimental data or from 
{\it ab initio} lattice QCD simulations. Due to such acute paucity of data it becomes imperative to rely on some of 
the erstwhile phenomenological models, before our LO EFT analysis can be made viable to yield some qualitative 
insight. Thus, for example, here we rely on the $\Xi^- nn$ binding energy estimates from the Faddeev calculations 
provided by the potential model analyses of
Refs.~\cite{Carames:2012zz,Garcilazo:2015noa,Garcilazo:2016ylj,Garcilazo:2016ams,Filikhin:2017fog}. 

\subsection{Coupled STM Integral Equations}
\label{sec:2}
%%%%%%%%%%%%%%%%%%%%%%%%%%%%%%%%%%%%%%%%%%%%%%%%%%%%%%%%%%%%%%%%%%%%%%%%%%%%%%%%%%%%%%%%%%%%
\begin{figure}[tbp]
\resizebox{1.0\columnwidth}{!}{\includegraphics{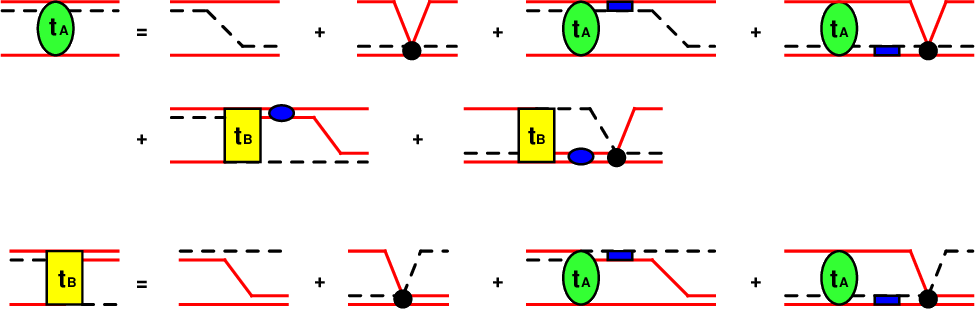}}
\caption{ Feynman diagrams for the representative coupled channel elastic scattering process, 
          $n+ (\Xi^- n)_t \rightarrow n + (\Xi^- n)_t$, where ``$t$" is used to denotes the ${}^3S_1$ $\Xi^- n$ 
          sub-system. The solid (dash) line represents neutron ($\Xi^-$-hyperon) propagator. The off-shell double 
          lines with insertions of the small empty oval (square) blobs represent the renormalized dressed 
          ${}^1S_0$ $nn$ ($u_0$) and ${}^3S_1$ $\Xi^- n$ ($u_1$) dibaryon field propagators. The large blob 
          $t_A$ ($t_B$) denotes the elastic (inelastic) half-off-shell scattering amplitude for the
          $n + u_1 \rightarrow n+ u_1$ ($n+ u_1 \rightarrow \Xi^- + u_0$) scattering processes. The dark blobs 
          represent the insertions of leading order three-body contact interactions.}
\label{fig-1}
\end{figure}
%%%%%%%%%%%%%%%%%%%%%%%%%%%%%%%%%%%%%%%%%%%%%%%%%%%%%%%%%%%%%%%%%%%%%%%%%%%%%%%%%%%%%%%%%%%%%
For the sake of theoretical analysis, we study the $\Xi^- nn$ system in both the kinematical three-body bound and 
scattering domains. For this purpose we choose a representative elastic reaction channel corresponding to the 
low-energy $1+2\to 1+2$ scattering process with dominant S-wave contribution, namely, 
\begin{equation}
 n + (\Xi^- n)_t \longrightarrow  n+ (\Xi^- n)_t \,\,. 
\label{eq12}
\end{equation}
Here we must emphasize that this ``reference" scattering process is chosen solely for demonstrating our theoretical
methodology, irrespective of the infeasibility of performing such experiments at current facilities. The chosen 
reaction channel yields a set of two coupled {\it Fredholm-type} integral equations in the momentum space. While their 
eigenvalues yield all possible allowed trimer binding energies ($B_3$), the eigenvectors yield the scattering 
amplitudes of the coupled elastic and inelastic channels. In Fig.~\ref{fig-1}, we display the relevant Feynman diagrams 
for the above mentioned scattering process expressed in term of the two {\it half-off-shell\,} S-wave projected 
scattering amplitudes, namely, $t_A(k,p;E)$ representing the elastic process $n + u_1 \rightarrow n + u_1$, and 
$t_B(k,p)$ representing the inelastic process $n + u_1 \rightarrow \Xi^- + u_0$. Here $k=|{\bf k}|\,\, (p=|{\bf p}|)$ 
denotes the on-shell (off-shell) incoming (outgoing) relative three-momentum in the center-of-mass (CM) system, and $E$ 
is the total CM kinetic energy of the three-body system, given by
\begin{equation}
 E=\pm\frac{k^2}{2\mu_{n(n\Xi)}} - \mathcal{B}_2\quad;\quad \mathcal{B}_2 
 = \frac{\left(\gamma^{(1)}_{\Xi n}\right)^2}{2 \mu}=1.47~{\rm MeV}\,,
\end{equation}
where the ``$-$'' sign is applicable for the kinematical three-body bound state domain and the ``$+$'' sign for the 
scattering domain. $\mathcal{B}_2$ is the CM binding energy of a possible shallow-bound $(\Xi^- n)_t$ sub-system which
also sets the scale for the particle-triplet-dimer ($n + u_1$) break-up threshold energy,\footnote{In our halo EFT 
formalism  ${\mathcal B}_2$ corresponds to the pole position of the $u_1$ dibaryon propagator. Its value may be 
compared with the binding energy of the putative $D^*$ state in the (1,1) $\Xi N$ channel, as predicted by the potential
model analyses of Refs.~\cite{Garcilazo:2016ams,Filikhin:2017fog,Nagels:2015dia,Rijken:2016uon}.} and 
$\mu_{n(n\Xi)}= M_n(M_{\Xi}+M_n)/(2M_n+M_{\Xi})=663.768$~MeV is the corresponding reduced mass of three-body 
(particle-dimer) system. The construction methodology of the integral equations is similar to those employed in several
earlier ${}^{{\pi}\!\!\!/}$EFT works~\cite{Raha:2017ahu,Ando:2015fsa,Ando:2013kba,Ando:2014mqa,Meher:2020rsd} (also see 
Appendix B). The renormalized S-wave projected coupled (elastic and inelastic channels) STM integral equations with the 
introduced cut-off regulator $\Lambda_{\rm reg}$ are given as
\begin{eqnarray}
t^{(R)}_A(p,k;E) &=&  {\mathcal Z}_{\Xi n}\frac{\left(y_1^2M_{\Xi} \right)}{2}
\left[K_{(a)}(p,k;E)-\frac{g_3(\Lambda_{\rm reg})}{\Lambda_{\rm reg}^2}\right]
\nonumber\\
&& -\, \frac{M_{\Xi}}{2\pi\mu}\!\int_0^{\Lambda_{\rm reg}} dq\,  q^2
\left[K_{(a)}(p,q;E)-\frac{g_3(\Lambda_{\rm reg})}{\Lambda_{\rm reg}^2}\right]
\nonumber\\
&& \hspace{4cm} \times \,{\mathscr D}_1\left(E-\frac{q^2}{2M_n},{\bf q}\right)t^{(R)}_A(q,k;E)
\nonumber\\
&& +\, \frac{\sqrt{6}y_1}{\pi y_0}\int_0^{\Lambda_{\rm reg}} dq\,  q^2
\left[K_{(b2)}(p,q;E)-\frac{g_3(\Lambda_{\rm reg})}{\Lambda_{\rm reg}^2}\right] 
\nonumber\\
&& \hspace{4cm} \times \, {\mathscr D}_0\left(E-\frac{q^2}{2M_\Xi},{\bf q}\right)t^{(R)}_B(q,k;E)\,\,,
\nonumber\\
\label{eq:11}
\end{eqnarray}
and
\begin{eqnarray}
t^{(R)}_B(p,k;E) &=& -{\mathcal Z}_{\Xi n}\sqrt{\frac{3}{2}} \left(y_1 y_0 M_n\right)
\left[K_{(b1)}(p,k;E)-\frac{g_3(\Lambda_{\rm reg})}{\Lambda_{\rm reg}^2}\right]
\nonumber\\
&& +\, \sqrt{\frac{3}{2}} \frac{M_n y_0}{\mu \pi y_1} \int_0^{\Lambda_{\rm reg}} dq\,  q^2
\left[K_{(b1)}(p,q;E)-\frac{g_3(\Lambda_{\rm reg})}{\Lambda_{\rm reg}^2}\right]
\nonumber\\
&& \hspace{4cm} \times \,{\mathscr D}_1\left(E-\frac{q^2}{2M_n},{\bf q}\right)t^{(R)}_A(q,k;E)\,\,,
\nonumber \\
\label{eq:12}
\end{eqnarray}
where the term $K_{(a)}$ denotes the S-wave projected $\Xi$-exchange interaction kernel, while $K_{(b_1,b_2)}$ 
are two different variants of the $n$-exchange interaction kernel, namely,
\begin{eqnarray}
  K_{(a)}   (p,\kappa;E) &=& \frac{1}{2p\kappa} \ln\left(\frac{p^2+\kappa^2
  +ap\kappa -2\mu E}{p^2+\kappa^2-ap\kappa -2\mu E}\right) \,\,,
  \nonumber\\
  \nonumber\\
  K_{(b_1)} (p,\kappa;E) &=&  \frac{1}{2p\kappa} \ln\left(\frac{bp^2+\kappa^2
  +p\kappa -M_n E}{bp^2+\kappa^2-p\kappa -M_n E}\right)\,\,, 
  \nonumber\\
  \nonumber\\
  K_{(b_2)} (p,\kappa;E) &=&  \frac{1}{2p\kappa} \ln\left(\frac{p^2
  +b\kappa^2+p\kappa -M_n E}{p^2+b\kappa^2-p\kappa -M_n E}\right)\,\,.
\end{eqnarray}
The generic momentum $\kappa$ denotes either the incoming on-shell relative momentum ($k$) or the loop 
momentum ($q$). Also, $a=2\mu/M_{\Xi}$ and $b=M_n/(2\mu)$ are two mass-dependent parameters. The above 
half-off-shell renormalized amplitudes are related to the corresponding unrenormalized amplitudes 
$t_{A,B}(p,k;E)$ by 
\begin{equation}
t^{(R)}_{A,B}(p,k;E)= \sqrt{{\mathcal Z}_{\Xi n}}\, t_{A,B}(p,k;E)\, \sqrt{{\mathcal Z}_{\Xi n}}\,\,,
\end{equation}
where
\begin{equation}
{\mathcal Z}^{-1}_{\Xi n}=\frac{y_1^2\mu^2}{2\pi \gamma^{(1)}_{\Xi n}}\,,
\label{EQWFZ}
\end{equation}
is the wavefunction renormalization associated with the possible bound $(\Xi^- n)_t$ sub-system. Finally, 
the renormalized elastic amplitude is used to obtain the S-wave $n-(\Xi^- n)_t$ three-body scattering length by 
considering the threshold limit of the on-shell momentum $k\to 0$, namely,
\begin{eqnarray}
\label{eq:a3}
  a_3=-\lim_{k\to 0}\frac{\mu_{n(n\Xi)}}{2\pi}\, t^{(R)}_A(k,k)\,.
\end{eqnarray}
 
\subsection{Asymptotic Analysis}
In order to assess that the coupled STM integral equations indeed have the potentiality to yield three-body 
bound state solutions, one needs to check for possible manifestation of Efimov effect at the asymptotic UV limit
as $\Lambda_{\rm reg} \rightarrow \infty$~\cite{Braaten:2004rn} (cf. Appendix A.3). In this case, all other 
low-energy/momentum scales in the problem, e.g., 
$E,\gamma^{(1)}_{\Xi n},\gamma^{(0)}_{nn},k \ll p,q\sim \Lambda_{\rm reg}\lesssim\infty$, become irrelevant, and 
the integral equations can be well approximated by considering only the homogeneous parts (i.e., excluding the 
tree diagram contributions in Fig.~\ref{fig-1}) and dropping all the $k$ dependence and three-body interactions 
($g_3$) terms. Thus, with no other relevant scales in the theory, the STM equations become {\it dilation} invariant 
and symmetric under the inversion transformation $q \to 1/q$. Consequently, the half-off-shell channel amplitudes 
exhibit a power-law scaling, namely, $t_{A,B}(\kappa)\sim \kappa^{s-1}$, with $\kappa\sim\Lambda_{\rm reg}$ and a complex-valued
exponent $s$, an undetermined three-body parameter. By performing a sequence of {\it Mellin} transformations the 
integral equations can be converted into a single transcendental equation which solves for the exponent $s$, namely, 
\begin{eqnarray}
  1 &=& \frac{M_{\Xi}}{2\mu C_1} \Bigg[ \frac{\sin[s \sin^{-1}(a/2)]}{s\cos(\pi s/2)}\Bigg] 
  + \frac{3M_n}{\mu C_1C_2} \Bigg[\frac{\sin[s \cot^{-1}\sqrt{4b-1}]}{s\cos(\pi s/2)}\Bigg]^2,
  \label{eq18}
\end{eqnarray}
where 
\begin{eqnarray}
  C_1=\sqrt{\frac{\mu}{\mu_{n(n\Xi)}}}, \quad{\text{and}}\quad 
  C_2=\sqrt{\frac{M_n}{2\mu_{\Xi(nn)}}}\,\,,
  \label{eq19}
  \end{eqnarray}
and $\mu_{\Xi(nn)} = 2 M_n M_{\Xi}/(2 M_n +M_{\Xi})=775.942$~MeV is the reduced mass of the $\Xi^- + u_0$ particle-dimer 
system. Solving Eq.~\eqref{eq18} yields an imaginary solution, i.e., $s=\pm i s_0^{\infty}= \pm i 0.803391\cdots $. The 
solution immediately suggests the existence of an asymptotic UV RG limit cycle with a discrete scaling symmetry associated
with the scale factor, $\lambda_\infty =e^{\pi/s^{\infty}_0}=49.919712\cdots$. This formally implies that our LO EFT 
manifests Efimov effect in the unitary limit of the $\Xi^- nn$ system. Consequently, it becomes imperative to include scale
dependent 3BF as counterterms in the effective Lagrangian to renormalize the ill-defined asymptotic limit of the STM 
equations with two-body interaction. As elucidated by power counting arguments in the Appendix A.2, such non-derivative 3BF
terms are naturally enhanced to get promoted to LO for consistency of the renormalization 
scheme~\cite{Bedaque:1998kg,Bedaque:1999ve}. Here we must, however, mention that the asymptotic scaling exponent 
$s^\infty_0=0.803391\cdots$ considerably differs from the expected value, $(s^\infty_0)_{\rm expect}\sim 1.01$, based on 
the universal RG limit cycle scaling depending on the relative three-particle mass ratios~\cite{Braaten:2004rn}. This 
difference is attributed to the effect of excluding the isospin-triplet spin-singlet (1,0) $\Xi^- n$ sub-system channel from
the STM equations whose dynamics are not directly amenable to our low-energy EFT description. 

%%%%%%%%%%%%%%%%%%%%%%%%%%%%%%%%%%%%%%%%%%%%%%%%%%%%%%%%%%%%%%%%%%%%%%%%%%%%%%%%%%%%%%%%%%%%%%%%%%%%%%%%%%%%%%%%%%%%
\section{Results and Discussion}\label{sec:3}
\begin{figure}[tbp]
\centering
\includegraphics[width=10cm]{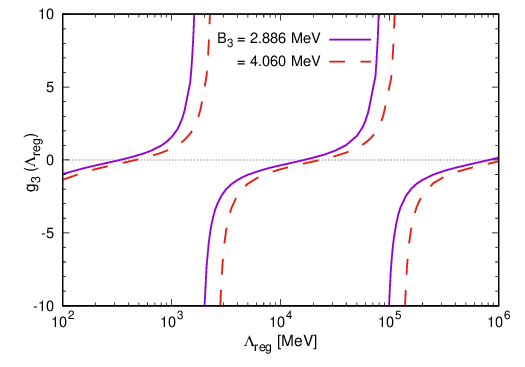}
\caption{ The approximate RG limit cycle behavior of the three-body coupling $g_3$ for the $\Xi^- nn$ ($I=3/2,\, J={1/2}$) 
          system as a function of the cut-off scale $\Lambda_{\rm reg}$. The results are obtained by numerically solving the STM
          integral equations~\eqref{eq:11} and \eqref{eq:12}. The input three-body binding energies $B_3= 2.886,\, 4.06$~MeV, 
          are predictions from the Faddeev calculation based potential models~\cite{Garcilazo:2015noa,Filikhin:2017fog}. 
          The input S-wave spin-isospin triplet $\Xi^- n$ scattering length $a^{(j=1)}_{\Xi n}=4.911$~fm is provided by the recent
          ESC08c Nijmegen potential model analyses~\cite{Nagels:2015dia,Rijken:2016uon}.}
\label{fig-2}
\end{figure}

%%%%%%%%%%%%%%%%%%%%%%%%%%%%%%%%%%%%%%%%%%%%%%%%%%%%%%%%%%%%%%%%%%%%%%%%%%%%%%%%%%%%%%%%%%%%%%%%%%%%%%%%%%%%%%%%%%%%
In this section we present the results of our preliminary investigation of the sharp cut-off regulator ($\Lambda_{\rm reg}$) 
dependence of the Faddeev-type STM integral equations~\eqref{eq:11} and \eqref{eq:12}, at non-asymptotic low-energy 
scales. For the sake of numerical evaluations, we use the particle masses as presented in Table~\ref{tab:1}, while the 
S-wave scattering lengths $a_{nn}=-18.63$~fm~\cite{Chen:2008zzj} and 
$a^{(j=1)}_{\Xi n}=4.911$~fm~\cite{Nagels:2015dia,Rijken:2016uon} constitute the principal input two-body parameters 
in our LO EFT framework. In the last section, our asymptotic analysis demonstrated the evidence of Efimov effect at the 
unitary limit of the $\Xi^- nn$ system with an RG limit cycle discrete scaling symmetry determined by the multiplicative
factor $\lambda_\infty=e^{\pi/s^\infty_0}\sim 50$. With $\kappa^{(1)}\equiv \gamma^{(1)}_{\Xi n}\sim 40$~MeV as the 
typical momentum scale of the problem, it is natural to expect that the next higher momentum scale appears at 
$\kappa^{(2)}\equiv \lambda_\infty \gamma^{(1)}_{\Xi n}\sim 2~{\rm GeV}\,\gg \Lambda_H\sim m_\pi$, which is well beyond 
the accessibility of our low-energy EFT description. Hence, it is likely that at the most one Efimov-like state emerges 
as a plausible bound $\Xi^- nn$ hypernucleus, if at all. Figure~\ref{fig-2} shows the cut-off dependence of the three-body 
contact interaction coupling $g_3(\Lambda_{\rm reg})$. In the absence of datum to constrain the unknown coupling $g_3$, 
our {\it strategy} is to hypothetically assume at the very outset that the $\Xi^- nn$ system is bound, with ground state 
eigenenergy $(E=-B_3)$ coinciding with the existing (Faddeev calculations) model predictions of
Refs.~\cite{Garcilazo:2015noa,Garcilazo:2016ylj,Garcilazo:2016ams,Filikhin:2017fog}. We thereby fix our benchmark range 
of input values of the $\Xi^-nn$ binding energy, namely, between $B_3= 2.886$~MeV taken from 
Ref.~\cite{Filikhin:2017fog}, and $B_3= 4.06$~MeV taken from Ref.~\cite{Garcilazo:2015noa}. Notably, both predictions 
rely on the same two-body input parameters (e.g., $a^{(1)}_{\Xi n}=4.911$~fm) provided by the recent ESC08c Nijmegen model 
analyses~\cite{Nagels:2015dia,Rijken:2016uon}.\footnote{The predicted value $B_3= 2.886$~MeV obtained in Faddeev 
calculation analysis of Ref.~\cite{Filikhin:2017fog} resulted from considering both the repulsive (1,0) and attractive 
(1,1) $\Xi N$ channels. Whereas, the value $B_3= 4.06$~MeV obtained in the Faddeev analysis of 
Ref.~\cite{Garcilazo:2015noa} resulted from considering only the latter attractive channel. Nevertheless, irrespective of these
details, we consider these predicted values as given three-body inputs to our EFT analysis.} The figure displays the typical
quasi-periodic log-singularities of {\it approximate} RG limit cycles for the two aforementioned limiting $B_3$ inputs. The 
corresponding non-asymptotic scale factor, $\lambda_n \lesssim \lambda_\infty$, may be obtained by considering the ratio of 
two successive cut-offs where the three-body coupling vanishes, i.e., if 
$g_3\left(\Lambda^{(n)}_{\rm reg}\right)=g_3\left(\Lambda^{(n+1)}_{\rm reg}\right)=0$, then  
%%%%%%%%%%%%%%%%%%%%%%%%%%%%%%%%%%%%%%%%%%%%%%%%%%%%%%%%%%%%%%%%%%%%%%%%%%%%%%%%%%%%%%%%%%%%
%%%%%%%%%%%%%%%%%%%%%%%%%%%%%%%%%%%%%%%%%%%%%%%%%%%%%%%%%%%%%%%%%%%%%%%%%%%%%%%%%%%%%%%%%%%%
\begin{equation}
 \lambda_{n} = \frac{\Lambda_{\rm reg}^{(n+1)}}{\Lambda_{\rm reg}^{(n)}}\,\,; \quad n = 1,2,..,\infty\,\,, 
\end{equation}
where $\Lambda_{\rm reg}^{(n)}$ is the cut-off corresponding to $n^{\rm th}$ zero of $g_3$. Table~\ref{tab:2} displays some
of the estimated non-asymptotic scale factors $\lambda_n$ corresponding to successive pairs of zeros of each RG limit cycle
obtained for the two limiting input $B_3$ values. In each case the scale factor $\lambda_n$ is obtained close to yet less 
than the asymptotic value $\lambda_\infty$. By progressively choosing larger pairs of the successive zeros of $g_3$, i.e., 
with $n \to \infty$, $\lambda_n$ is found to converge rapidly to $\lambda_\infty$.  

%%%%%%%%%%%%%%%%%%%%%%%%%%%%%%%%%%%%%%%%%%%%%%%%%%%%%%%%%%%%%%%%%%%%%%%%%%%%%%%%%%%%%%%%%%%%
\begin{table}[tbp]
\begin{center}
\caption{The approximate RG limit cycle behavior with the discrete scaling symmetry factor $\lambda_n\to \lambda_\infty$, 
         obtained by solving the integral equations~\eqref{eq:11} and \eqref{eq:12} for the $\Xi^- nn$ ($I=3/2,\, J={1/2}$) 
         system. Here, results for $n\leq 4$ display a rapid convergence of the scale parameter toward the asymptotic limit,
         $\lambda_\infty=49.919712\cdots$. The input three-body binding energies $B_3= 2.886,\,4.06$~MeV are predictions 
         from the Faddeev calculation based potential models~\cite{Garcilazo:2015noa,Filikhin:2017fog} with input S-wave 
         $\Xi^- n$ ${}^3S_1$ scattering length $a^{(1)}_{\Xi n}=4.911$~fm, provided by the ESC08c Nijmegen potential model 
         analyses~\cite{Nagels:2015dia,Rijken:2016uon}. } 
\label{tab:2}  
\begin{tabular}{|c|c|c|c|c|}
\hline
Binding Energy & $n\in{\mathbb Z}_+$ & $n^{\rm th}$ zero of $g_3$ & $(n+1)^{\rm th}$ zero of $g_3$ & Scale factor\\
  $B_3$ (MeV)  &                   & $\Lambda_{\rm reg}^{(n)}$ (MeV)    & $\Lambda_{\rm reg}^{(n+1)}$ (MeV)      & $\lambda_n=\Lambda^{(n+1)}/\Lambda^{(n)}$ \\
\hline\hline
                                & 1 & 334.283      & 16344.134      & $48.893105\cdots$ \\ 
2.886 \cite{Filikhin:2017fog}   & 2 & 16344.134    & 815412.631     & $49.890232\cdots$ \\ 
                                & 3 & 815412.631   & 40704680.527   & $49.919119\cdots$ \\ 
                                & 4 & 40704680.527 & 2031965537.021 & $49.919702\cdots$ \\ 
\hline
                                & 1 & 465.937      & 22919.007      & $49.189069\cdots$ \\
4.06 \cite{Garcilazo:2015noa}   & 2 & 22919.007    & 1143628.429    & $49.898690\cdots$ \\
                                & 3 & 1143628.429  & 57089119.370   & $49.919290\cdots$ \\
                                & 4 & 57089119.370 & 2849872042.899 & $49.919706\cdots$ \\
\hline\hline
\end{tabular}
\end{center}
\end{table}
%%%%%%%%%%%%%%%%%%%%%%%%%%%%%%%%%%%%%%%%%%%%%%%%%%%%%%%%%%%%%%%%%%%%%%%%%%%%%%%%%%%%%%%%%%%%

\vspace{0.1cm}

\begin{figure}[tbp]
\hspace{-0.5cm}
\includegraphics[width=6.59cm]{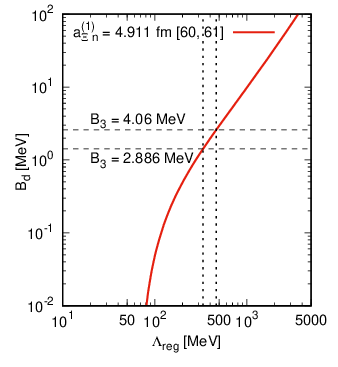}~\includegraphics[width=6.59cm]{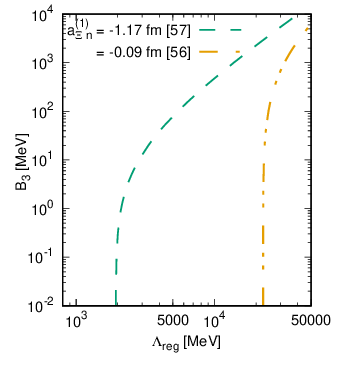}
    \caption{Cut-off regulator ($\Lambda_{\rm reg}$) dependence of the three-body binding energy of the $\Xi^- nn$ 
             ($I=3/2,\, J={1/2}$) system, obtained by solving the coupled integral equations~\eqref{eq:11} and \eqref{eq:12}, 
             excluding the three-body contact interactions [i.e. $g_3(\Lambda_{reg})=0$]. {\bf Left panel:} Three-body binding energy $B_d=B_3-{\mathcal B}_2$, 
             relative to the $n+(\Xi^-n)_t$ particle-dimer threshold $-E={\mathcal B}_2=1.47$~MeV, with the input S-wave 
             ${}^3S_1$ $\Xi^- n$ scattering length $a^{(1)}_{\Xi n}=4.911$~fm, as predicted by the recently updated ESC08c Nijmegen potential 
             model analyses~\cite{Nagels:2015dia,Rijken:2016uon}. The regulator independent predictions, namely, 
             $B_3= 2.886$~MeV and $4.06$~MeV, from the Faddeev calculation based potential model 
             analyses~\cite{Garcilazo:2015noa,Filikhin:2017fog} for the same $a^{(j=1)}_{\Xi n}$ input are displayed for 
             comparison. {\bf Right panel:} Three-body binding energy $B_3$ relative to the three-particle threshold with input 
             $a^{(j=1)}_{\Xi n}=-0.09\,,-1.17$~fm, as predicted by the two recent SU(3) chiral EFT 
             analyses~\cite{Li:2018tbt,Haidenbauer:2018gvg}. }
\label{fig-3}
\end{figure}

Next, in Fig.~\ref{fig-3} we display the cut-off variation of the binding energy $B_3$ excluding the 3BF terms, i.e., with 
$g_3=0$ in the STM equations. In particular, due to the ambiguities concerning the precise nature of the (1,1) $\Xi N$
sub-system interactions between different existing phenomenological
analyses~\cite{Khaustov:1999bz,Hiyama:2019kpw,Aoki:1998sv,Tamagawa:2001tk,Ahn:2005jz,Yamaguchi:2001ip,Friedman:2007zza,Hiyama:2010zz,Haidenbauer:2015zqb,Li:2018tbt,Haidenbauer:2018gvg,Kohno:2009ny,Krishichayan:2010zza,Nagels:2015dia,Rijken:2016uon}, we consider here two representative scenarios with contrasting 
perspectives as elucidated below (both cases can formally lead to the emergence of Efimov-like states): 
\begin{itemize}
\item 
First, the scenario with the input positive $(\Xi^-n)_t$ scattering length, e.g., $a^{(j=1)}_{\Xi n}=4.911$~fm, as predicted 
by the updated ESC08c Nijmegen potential model analyses of Refs.~\cite{Nagels:2015dia,Rijken:2016uon}, suggests a strongly 
attractive ${}^3S_1$ $\Xi^- n$ sub-system commensurate with the likely existence of a threshold bound state
($D^*$)~\cite{Garcilazo:2016ams,Filikhin:2017fog}. Consequently, in the three-body sector with a pair of likely bound 
$(\Xi^- n)_t$ sub-systems and a virtual bound $nn$ sub-system, the $\Xi^- nn$ system assumes a halo-bound 
{\it samba-configuration}~\cite{Yamashita:2004pv} structure emerging from the particle-triplet-dimer ($n+(\Xi^- n)_t$) 
break-up threshold at the CM energy $E=-{\mathcal B}_2$. This corresponds to the solid (red) line curve in the left panel of
Fig.~\ref{fig-3}, representing the regulator dependence of the {\it relative} binding energy $B_d=B_3-{\mathcal B}_2$, for the
ground ($n=0$) Efimov-like state which appears at the {\it critical cut-off} scale $\Lambda^{(0)}_{\rm crit}\approx 80$~MeV. \\
\item 
Second, the scenario with the input negative $(\Xi^-n)_t$ scattering length, e.g., as predicted by the two recent SU(3) chiral
EFT analyses, namely, $a^{(j=1)}_{\Xi n}=-0.09$~fm~\cite{Li:2018tbt} and $-1.17$~fm~\cite{Haidenbauer:2018gvg}, suggests a 
weakly attractive ${}^3S_1$ $\Xi^- n$ sub-system that is unlikely to exhibit any two-body bound state. Consequently, in the 
three-body sector with no bound two-body sub-systems, the $\Xi^- nn$ system assumes a bound 
{\it borromean-configuration}~\cite{Yamashita:2004pv} structure emerging from the three-particle break-up threshold $E=0$. This
corresponds to the two broken line curves in the right panel of Fig.~\ref{fig-3}, representing the regulator dependence of $B_3$
for the respective ground Efimov-like states which appear above the threshold at the critical values, 
$\Lambda^{(0)}_{\rm crit} \approx 1940$~MeV for Ref.~\cite{Haidenbauer:2018gvg} and $22470$~MeV for Ref.~\cite{Li:2018tbt}.
\end{itemize}
Evidently, with such large critical cut-offs, the latter scenario most likely not be supported in our low-energy EFT framework, 
{\it vis-a-vis}, the Efimov-like ground state does not physically manifest as a bound $\Xi$-hypernucleus. In contrast, the
small critical cut-off in the former scenario lies well within the EFT validity domain, indicating an encouraging prospect 
for a potentially feasible $\Xi^- nn$ Efimov state. In what follows, we shall only discuss our results pertaining to the 
former choice of the $\Xi ^- nn$ scenario.
%%%%%%%%%%%%%%%%%%%%%%%%%%%%%%%%%%%%%%%%%%%%%%%%%%%%%%%%%%%%%%%%%%%%%%%%%%%%%%%%%%%%%%%%%%%%
%%%%%%%%%%%%%%%%%%%%%%%%%%%%%%%%%%%%%%%%%%%%%%%%%%%%%%%%%%%%%%%%%%%%%%%%%%%%%%%%%%%%%%%%%%%%
%%%%%%%%%%%%%%%%%%%%%%%%%%%%%%%%%%%%%%%%%%%%%%%%%%%%%%%%%%%%%%%%%%%%%%%%%%%%%%%%%%%%%%%%%%%%
\begin{figure}[h]
\centering
\includegraphics[width=10cm]{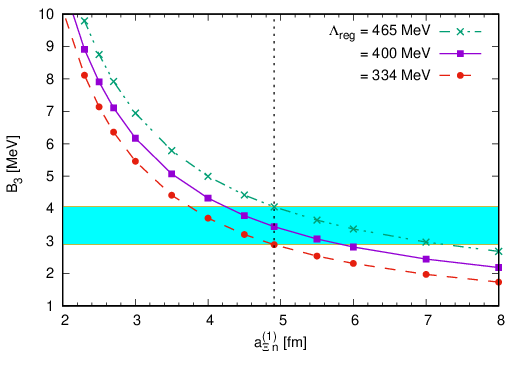}
    \caption{Variation of the three-body binding energy $B_3$ of the $\Xi^- nn$ ($I=3/2,\, J={1/2}$) system as a function 
             of input positive values of the S-wave ${}^3S_1$ $\Xi^- n$ scattering length $a^{(1)}_{\Xi n}$ for fixed cut-offs 
             $\Lambda_{\rm reg}$ excluding three-body interactions. The horizontal shaded band represents our benchmark range 
             of values of $B_3$ considered between the limits, $B_3=2.886$~MeV and $4.06$~MeV, predicted by the Faddeev 
             calculation based potential model analyses~\cite{Garcilazo:2015noa,Filikhin:2017fog}. The vertical dotted line 
             represents our choice of the input scattering length $a^{(1)}_{\Xi n} = 4.911$~fm, as predicted by the recently updated 
             ESC08c Nijmegen potential model analyses~\cite{Nagels:2015dia,Rijken:2016uon}.   }
\label{fig-4} 
\end{figure}
%%%%%%%%%%%%%%%%%%%%%%%%%%%%%%%%%%%%%%%%%%%%%%%%%%%%%%%%%%%%%%%%%%%%%%%%%%%%%%%%%%%%%%%%%%%%

\vspace{-0.3cm}

In the absence of the three-body contact interactions for renormalization our results for the three-body binding energy 
exhibit considerable sensitivity to the cut-off variations. Figure~\ref{fig-3} also compares our results with the regulator
independent predictions for the $\Xi^- nn$ binding energy from the potential models~\cite{Garcilazo:2015noa,Filikhin:2017fog} 
which also rely on the two-body inputs from the Nijmegen ESC08c model analyses~\cite{Nagels:2015dia,Rijken:2016uon}. We find 
that our scale dependent eigenenergies from the STM equations reproduce the model predictions, namely, $B_3=2.886$~MeV of 
Ref.~\cite{Filikhin:2017fog} and $B_3=4.06$~MeV of Ref.~\cite{Garcilazo:2015noa} at the cut-off scales 
$\Lambda_{\rm reg}\approx 334$~MeV and $\Lambda_{\rm reg}\approx 465$~MeV respectively. The same result is demonstrated more 
conspicuously in Fig.~\ref{fig-4} where we plot the variation of the eigenenergy $B_3$ by (hypothetically) varying the 
scattering length $a^{(1)}_{\Xi n}>0$ for several fixed cut-offs $\Lambda_{\rm reg}$ excluding 3BF terms. The chosen potential model 
predicted range, $2.886~{\rm MeV} \lesssim B_3\lesssim 4.06~{\rm MeV}$, as demarcated by the horizontal band
in the figure, is well constrained within our regulator range, 
$334~{\rm MeV} \lesssim \Lambda_{\rm reg} \lesssim 465~{\rm MeV}$. In particular, our summary Table~\ref{tab:3} displays the 
$\Lambda_{\rm reg}$ values at which our EFT solutions reproduce several more of the existing Faddeev calculation based model 
predictions for the $\Xi^- nn$ binding energy~\cite{Garcilazo:2015noa,Garcilazo:2016ylj,Garcilazo:2016ams,Filikhin:2017fog}. 
Although the above regulator range apparently seems well beyond the expected ${}^{{\pi}\!\!\!/}$EFT hard scale 
$\Lambda_H\sim m_\pi$, the model results may still be accommodated within the framework of a modified EFT having an 
extended domain of validity. Consequently, such a modified halo ${}^{{\pi}\!\!\!/}$EFT should have a larger breakdown scale, 
say, $\widetilde{\Lambda}_H\lesssim 500$~MeV, where interactions between the $\Xi$-hyperon and neutron are possibly dominated by 
two-pion ($\pi\pi$) or $\sigma$-meson exchange mechanisms. We note that one-pion-exchanges are typically ruled out by isospin 
invariance in strong processes.  

\vspace{0.1cm}

\begin{figure}[tbp]
\centering
\includegraphics[width=6.58cm]{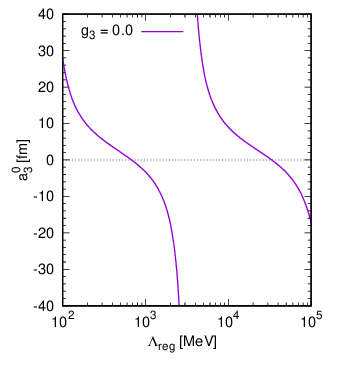}~\includegraphics[width=6.58cm]{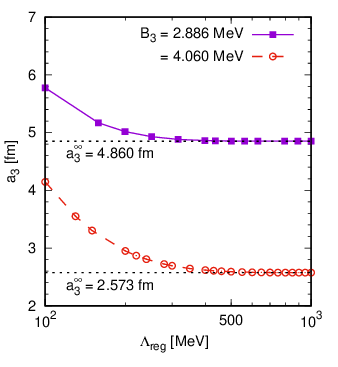}
    \caption{Regulator ($\Lambda_{\rm reg}$) dependence of the  $n-(\Xi^-n)_t$ elastic S-wave three-body scattering length $a_{3}$, 
             obtained by solving the coupled integral equations~\eqref{eq:11} and \eqref{eq:12} with input S-wave scattering length 
             $a_{\Xi n}=4.911$~fm, taken from the updated Nijmegen model analyses~\cite{Nagels:2015dia,Rijken:2016uon}. {\bf Left panel:} 
             The unrenormalized scattering length $a_3\to a^{0}_3$ excluding the three-body coupling, i.e., $g_3=0$. {\bf Right panel:} 
             The renormalized scattering length including the three-body coupling $g_3\neq 0$. The scale dependence of 
             $g_3(\Lambda_{\rm reg})$ is determined using the respective RG limit cycles (cf. Fig.~\ref{fig-2}) corresponding to the 
             two three-body inputs, $B_3= 2.886$~MeV and $4.06$~MeV, taken from the Faddeev calculation model 
             analyses~\cite{Garcilazo:2015noa,Filikhin:2017fog}. Our predictions, namely, $a_{3}^{\infty}= 4.860$~fm and $2.573$~fm, 
             correspond to the respective asymptotic limits. }
\label{fig-5} 
\end{figure}

Finally, we give a simple demonstration of predictability of our EFT framework. To this end, we attempt a {\it naive} 
estimation of the $\Xi^- nn$ ($I=3/2,\, J={1/2}$) three-body scattering length, or more precisely the $n-(\Xi^- n)_t$ 
elastic S-wave scattering length $a_3$, by utilizing the potential model predicted three-body binding energy information 
from Refs.~\cite{Garcilazo:2015noa,Garcilazo:2016ylj,Garcilazo:2016ams,Filikhin:2017fog}. Here we need to solve our 
coupled STM integral equations~\eqref{eq:11} and \eqref{eq:12} in the kinematical scattering domain. Subsequently, the 
three-body scattering length is obtained by considering the on-shell threshold limit of the renormalized elastic
scattering amplitude $t^{(R)}_A$ [cf. Eq.~\eqref{eq:a3}]. Solving the STM equations with the 3BF terms excluded (i.e., 
with $g_3=0$) leads to strong regulator dependence with the resulting amplitude displaying quasi-periodic singularities
akin to the limit cycle behavior (cf. left panel of Fig.~\ref{fig-5}). Such divergences are renormalized by introducing 
the 3BF counterterms with the running coupling $g_3(\Lambda_{\rm reg})$ already fixed using the RG limit cycles 
corresponding to model predicted $B_3$ inputs (cf. Fig.~\ref{fig-2}).
%%%%%%%%%%%%%%%%%%%%%%%%%%%%%%%%%%%%%%%%%%%%%%%%%%%%%%%%%%%%%%%%%%%%%%%%%%%%%%%%%%%%%%%%%%%%
%%%%%%%%%%%%%%%%%%%%%%%%%%%%%%%%%%%%%%%%%%%%%%%%%%%%%%%%%%%%%%%%%%%%%%%%%%%%%%%%%%%%%%%%%%%%

\vspace{0.1cm}

Figure~\ref{fig-5} (right panel) depicts the regulator dependence of the three-body scattering length 
$a_{3}(\Lambda_{\rm reg})$ renormalized by the 3BF terms. As mentioned, the scale dependence of the 3BF coupling 
$g_3(\Lambda_{\rm reg})$ is fixed using the RG limit cycles corresponding to the potential model inputs for 
$B_3$~\cite{Garcilazo:2015noa,Filikhin:2017fog}. The renormalized plots still exhibit a residual regulator dependence 
stemming from the low cut-off scale sensitivity of the counterterms owing to the decoupling of most underlying physics. 
However, for sufficiently large cut-off, say $\Lambda_{\rm reg}\gtrsim 400$~MeV, most of the underlying low-energy 
three-body dynamics are well captured in our solutions to the integral equations. Consequently, renormalizing $a_3$ 
using the counterterms becomes more effective at large $\Lambda_{\rm reg}$ leading to a well-defined asymptotic limit:
\begin{equation}
    a^\infty_3=\lim_{\Lambda_{\rm reg}\to \infty} a_{3}(\Lambda_{\rm reg}) \,.
\end{equation}
Hence, for each $B_3$ input a constant value $a^\infty_3$ is obtained, as demanded by renormalization invariance, representing our predicted three-body scattering length. In particular, our limiting benchmark inputs, $B_3=2.886$~MeV and 
$4.06$~MeV, lead to $a_{3}^{\infty}= 4.860$~fm and $2.573$~fm respectively. In addition, our summary Table~\ref{tab:3} 
displays some intermediate results corresponding to two other existing model predictions, namely, 
$B_3=3.89$~MeV and $3.00$~MeV, from the Faddeev calculations analyses of Refs.~\cite{Garcilazo:2016ylj,Garcilazo:2016ams}. 
Here we point out that for possible negative choice of the ${}^3S_1$ $\Xi^- n$ scattering length, such as the two recent 
SU(3) chiral EFT predictions, namely, $a^{(1)}_{\Xi n}=-0.09,\,-1.17$~fm~\cite{Li:2018tbt,Haidenbauer:2018gvg}, the 
$(\Xi^- n)_t$ sub-system is unbound with no kinematical particle-dimer scattering domain below the three-particle 
break-up threshold, i.e., $E<0$. 
%%%%%%%%%%%%%%%%%%%%%%%%%%%%%%%%%%%%%%%%%%%%%%%%%%%%%%%%%%%%%%%%%%%%%%%%%%%%%%%%%%%%%%%%%%%%
\begin{table}[tbp]
\begin{center}
\caption{Summary of our EFT results with three different input S-wave ${}^3S_1$ $\Xi^- n$ scattering lengths, namely, 
         $a^{(1)}_{\Xi n}=4.911$~fm, taken from the updated ESC08c Nijmegen model 
         analyses~\cite{Nagels:2015dia,Rijken:2016uon}, $a^{(1)}_{\Xi n}=-0.09$~fm, taken from the relativistic LO 
         chiral EFT analysis~\cite{Li:2018tbt}, and $a^{(1)}_{\Xi n}=-1.17$~fm, taken from the NLO chiral EFT-based 
         non-relativistic G-matrix analysis~\cite{Haidenbauer:2018gvg}. Displayed are the regulator scales 
         $\Lambda^{(g_3=0)}_{\rm reg}$ at which the Efimov ground state eigenenergy (by excluding $g_3$) reproduces 
         each of several existing potential model predictions on the three-body binding energies $B_3$ of the $\Xi^- nn$
         system~\cite{Garcilazo:2015noa,Garcilazo:2016ylj,Garcilazo:2016ams,Filikhin:2017fog}. Also summarized are our 
         predicted three-body scattering length ($a^\infty_3$) corresponding to each model input for $B_3$, with the 
         three-body coupling $g_3(\Lambda_{\rm reg})$ determined by the respective RG limit cycles. The results 
         corresponding to the $a^{(1)}_{\Xi n}<0$ scenario have no kinematical particle-dimer scattering domain for 
         $E<0$ and the three-body system is likely to remain unbound. In contrast, the $a^{(1)}_{\Xi n}>0$ scenario 
         shows encouraging prospect for a physically realizable $\Xi^-nn$ Efimov state. } 
\label{tab:3}  
\begin{tabular}{|c||c|c||c|}
\hline%\noalign{\smallskip}
Scattering                                          & Binding                        & Cut-off                   & Scattering \\
length $a^{(1)}_{\Xi n}$(fm)                        & energy $B_3$ (MeV)             & $\Lambda^{(g_3=0)}_{\rm reg}$ (MeV) & length $a^\infty_{3}$ (fm) \\
\hline\hline
                                                    & 2.886~\cite{Filikhin:2017fog}  & 334                       &  4.860 \\
      4.911                                         & 2.89~\cite{Garcilazo:2016ylj}  & 335                       &  4.849 \\
(Nijmegen model)~\cite{Nagels:2015dia,Rijken:2016uon}  & 3.00~\cite{Garcilazo:2016ams}  & 348                    &  4.562 \\
                                                    & 4.06~\cite{Garcilazo:2015noa}  & 465                       &  2.573 \\
\hline
                                                    & 2.886~\cite{Filikhin:2017fog}  & 22590                     &  - \\
      -0.09                                         & 2.89~\cite{Garcilazo:2016ylj}  & 22591                     &  - \\
(Relativistic Chiral EFT)~\cite{Li:2018tbt}         & 3.00~\cite{Garcilazo:2016ams}  & 22595                     &  - \\
                                                    & 4.06~\cite{Garcilazo:2015noa}  & 22633                     &  - \\
\hline
                                                    & 2.886~\cite{Filikhin:2017fog}  & 2333                      &  - \\
      -1.17                                         & 2.89~\cite{Garcilazo:2016ylj}  & 2334                      &  - \\
(G-matrix Chiral EFT)~\cite{Haidenbauer:2018gvg}       & 3.00~\cite{Garcilazo:2016ams}  & 2345                   &  - \\
                                                    & 4.06~\cite{Garcilazo:2015noa}  & 2440                      &  - \\
\hline\hline
\end{tabular}
\end{center}
\end{table}\\
%%%%%%%%%%%%%%%%%%%%%%%%%%%%%%%%%%%%%%%%%%%%%%%%%%%%%%%%%%%%%%%%%%%%%%%%%%%%%%%%%%%%%%%%%%%%

\vspace{-0.3cm}

\begin{figure}
\centering
    \includegraphics[width=10cm]{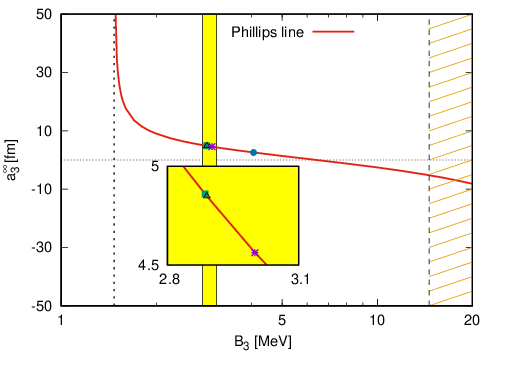}
    \caption{Phillips line correlation for the ($I=3/2,\,J=1/2$) $\Xi^-nn$ system corresponding to the input 
             ${}^3S_1$ $\Xi^- n$ scattering length $a^{(1)}_{\Xi n}=4.911$~fm, as predicted by the updated ESC08c Nijmegen 
             model analyses~\cite{Nagels:2015dia,Rijken:2016uon}. The data points correspond to the input values of the 
             three-body binding energy $B_3= 2.886~{\rm MeV},\, 2.89~{\rm MeV},\, 3.00~{\rm MeV}$ and $4.06$~MeV, 
             predicted by the potential model 
             analyses~\cite{Garcilazo:2015noa,Garcilazo:2016ylj,Garcilazo:2016ams,Filikhin:2017fog}. The vertical dotted 
             line on the left represents the $n+(\Xi^- n)_t$ particle-dimer threshold at $B_3=\mathcal{B}_2=1.47$~MeV, 
             while the hashed region, $B_3\gtrsim  14$~MeV, represents the expected breakdown region of our halo EFT 
             description. }
\label{fig-6} 
\end{figure}

Our predicted range of three-body scattering lengths represents a {\it naive} ballpark estimate based on the induced 
universal correlations which are expected to manifest in a halo-bound $\Xi^- nn$ system, albeit approximations 
considered in the analysis. To elucidate one of the inherent universal features, it is worth demonstrating the $B_3$
versus $a^\infty_3$ correlations corresponding to our preferred choice of the input $\Xi^- n$ scattering length, namely,
$a^{(1)}_{\Xi n}=4.911$~fm~\cite{Nagels:2015dia,Rijken:2016uon}. This yields the well-known Phillips line correlation 
plot for the $\Xi^- nn$ system, as depicted in Fig.~\ref{fig-6}. The curve diverges as $B_3\to {\mathcal B}_2=1.47$~MeV, 
the $n+(\Xi^- n)_t$ particle-triplet-dimer threshold, whenever an Efimov-like bound state emerges at zero-energy 
threshold (i.e., with $B_d=B_3-{\mathcal B}_2=0$). A second virtual bound three-body state with large negative value of 
$a^\infty_3$ is expected to emerge around $B^{\rm (virt)}_3=e^{\pi/s_0}{\mathcal B}_2\approx 70$~MeV, where the Phillips 
line diverges again. However, the latter state lies outside the domain of validity of standard ${}^{{\pi}\!\!\!/}$EFT 
with an estimated breakdown energy scale $-E_{\rm break}\sim 14$~MeV, as determined by the three-body binding momentum of 
the order of the pion mass, i.e., $\gamma_3=\sqrt{-2\mu_{n(n\Xi )}E_{\rm break}}\sim m_\pi$. The four data points 
displayed in the figure correspond to the $B_3$ predictions from the potential model 
analyses~\cite{Garcilazo:2015noa,Garcilazo:2016ylj,Garcilazo:2016ams,Filikhin:2017fog}, all of which rely on the same 
two-body input from the Nijmegen model analyses, namely,  
$a^{(1)}_{\Xi n}=4.911$~fm~\cite{Nagels:2015dia,Rijken:2016uon} (cf. Table~\ref{tab:3}). The fact that the Phillips plot 
evidently reflects certain degree of compatibility of the potential model inputs with our EFT description is an 
important qualitative finding of this work. In the event of possible future availability of phenomenological three-body 
data, a more rigorous NLO EFT analysis (explicitly including effective range terms) may be helpful to substantiate such 
connections on a better footing. 
%%%%%%%%%%%%%%%%%%%%%%%%%%%%%%%%%%%%%%%%%%%%%%%%%%%%%%%%%%%%%%%%%%%%%%%%%%%%%%%%%%%%%%%%%%%%
%%%%%%%%%%%%%%%%%%%%%%%%%%%%%%%%%%%%%%%%%%%%%%%%%%%%%%%%%%%%%%%%%%%%%%%%%%%%%%%%%%%%%%%%%%%%

%%%%%%%%%%%%%%%%%%%%%%%%%%%%%%%%%%%%%%%%%%%%%%%%%%%%%%%%%%%%%%%%%%%%%%%%%%%%%%%%%%%%%%%%%%%%%%%%%%%%%%%%%%%%%%%%%%%%
\section{Summary and Conclusions}\label{sec:4}
%%%%%%%%%%%%%%%%%%%%%%%%%%%%%%%%%%%%%%%%%%%%%%%%%%%%%%%%%%%%%%%%%%%%%%%%%%%%%%%%%%%%%%%%%%%%%%%%%%%%%%%%%%%%%%%%%%%%
A knowledge of few-body dynamics in light ($S=-2$) $\Xi$-hypernuclei can serve as an essential input to the neutron 
star EoS for possible explanation of their stabilities with masses $\gtrsim 2M_{\odot}$. In this regard, the 
$\Xi^- nn$ ($I=3/2,\, J^P={1/2}^+$) three-body system is one of the simplest systems to investigate the nature of the 
underlying 3BF. The reason being the stability of this channel against strong decays and the absence of Coulomb 
interactions. However, the impracticability of performing $\Xi$-hyperon scattering experiments and the lack of empirical 
data thereof have so far eluded rigorous determination of essential few-body observables. Thus, a general qualitative 
insight relying solely on low-energy universality is needed to illuminate specific characteristics of the underlying 
interactions that may reflect the emergence of exotic halo-bound states.

\vspace{0.1cm}

Here we used the framework of leading order halo ${}^{{\pi}\!\!\!/}$EFT in a speculative study to explore the feasibility 
that the $\Xi^- nn$ system is Efimov-bound. Notably, the presence of the predominantly repulsive ${}^1S_0$ $\Xi^- n$ 
sub-system channel potentially leads to the generation of anomalously deep two- and three-body unphysical bound states 
beyond the breakdown scale of the theory. In our current simplistic approach, as a first approximation, such unphysical 
contributions are avoided on an {\it ad hoc} basis by explicitly decoupling this channel in the construction of our 
integral equations. Our asymptotic analysis of the $\Xi^- nn$ integral equations revealed the formal appearance of Efimov 
states in the unitary limit associated with an RG limit cycle with a discrete scaling factor, $s_0^{\infty}=0.803391\cdots$.
The factor, however, differs from the expected value $(s_0^{\infty})_{\rm expect}\approx 1.01$, as governed by the 
universality of three-particle mass ratios~\cite{Braaten:2004rn}, owing to the decoupling of the ${}^1S_0$ $\Xi^- n$ 
channel. Such scaling differences can certainly influence various numerical estimates in the low-energy non-asymptotic 
domain, but the general qualitative features are likely to remain unchanged with the inclusion of both the spin 
channels.\footnote{In either cases we expect to find robust three-body universal features, such as the quasi-periodic RG 
limit cycle behavior of the three-body coupling and the induced Phillips line correlations.} 
%
%However, with the inclusion of the repulsive ${}^1S_0$ $\Xi^- n$ channel, we {\it naively} expect a shift of the RG limit cycles towards higher cut-off values which reduces the likelihood of a viable Efimov state. Nevertheless, this does not pre-empt the possibility that the anticipated scale shift gets compensated by the corresponding increase of scale factor $s^\infty_0$ to the expected value $\sim 1.01$.} 
%

\vspace{0.1cm}

Evidently, the unavailability of empirical three-body datum to fix the scale dependence of the three-body coupling 
$g_3(\Lambda_{\rm reg})$ is a major drawback of our approach which prevents robust predictions. Thus, we relied on 
several existing Faddeev calculation based potential model 
analyses~\cite{Garcilazo:2015noa,Garcilazo:2016ylj,Garcilazo:2016ams,Filikhin:2017fog} for the input three-body 
binding energy $B_3$ in the reasonable benchmark range, $2.886-4.06$~MeV. Moreover, the ${}^{{\pi}\!\!\!/}$EFT 
formalism requires the two-body inputs, namely, the ${{}^1S_0}$ $nn$ scattering length 
$a_{nn}=-18.63$~\cite{Zyla:2020zbs}, and the ${}^3S_1$ $\Xi^- n$ scattering length $a^{(1)}_{\Xi n}$. For the choice 
of the latter, we considered two contrasting scenarios, given the current ambiguities in regard to the underlying 
nature of the interactions in spin-isospin triplet (1,1) $\Xi^- n$ channel. In the first scenario we considered the 
prediction, $a_{\Xi n}^{(1)}=4.911$~fm, from the recently updated Nijmegen model analyses~\cite{Nagels:2015dia,Rijken:2016uon}, 
which is based on the notion of a strongly attractive likely bound $(\Xi^-n)_t$ sub-system. In the second 
scenario we considered the predictions of two contemporary SU(3) chiral EFT analyses, namely, 
$a_{\Xi n}^{(1)}=-0.09$~fm~\cite{Li:2018tbt}, based on a relativistic calculation, and 
$a_{\Xi n}^{(1)}=-1.17$~fm~\cite{Haidenbauer:2018gvg}, based on a non-relativistic in-medium G-matrix calculation, 
both concurring on a moderately attractive nature of the $(\Xi^-n)_t$ sub-system channel. With both the chiral EFT 
inputs, our investigations hinted at a predominantly unbound $\Xi^-nn$ system. In contrast with the input, 
$a_{\Xi n}^{(1)}=4.911$~fm, the first scenario indicated favourable prospects for a physically realizable $\Xi^-nn$ Efimov-like ground, 
with the proviso that our ${}^{{\pi}\!\!\!/}$EFT formalism could be extended (with 
$\widetilde{\Lambda}_H\gtrsim 500$~MeV) to accommodate interactions mediated by $\pi\pi$ or $\sigma$-meson exchanges. 
Specifically, our eigensolutions ($B_3$) to the integral equations (excluding the 3BF terms) could reproduce the 
benchmark range of model inputs for the cut-off regulator values in the range $\Lambda_{\rm reg}\approx 335-464$ MeV.  

\vspace{0.1cm}

Finally, as a simple demonstration of the predictive power of our EFT formalism we evaluated the three-body S-wave 
$n-(\Xi^- n)_t$ scattering length to lie in the range, $a^{\infty}_3\approx 2.6-4.9$~fm, corresponding to the same aforementioned 
benchmark range of model inputs for $B_3$. Given the very speculatory nature of the present study and the 
indeterminable three-body scattering length from present day experiments, the numerical figures for $a^\infty_3$ are 
by all means {\it naive} ballpark estimates. Nevertheless, they are indicative of the emergent universal features of 
a prospective Efimov-bound $\Xi^- nn$ system which induce three-body correlations like the Phillips line. Such 
universal three-body features are robust against ambiguities in the two-body description given that the 
three-body datum is reliable. Consequently, our obtained results reasonably guesstimate similar qualitative results 
expected from more rigorous ${}^{{\pi}\!\!\!/}$EFT-based future investigations with systematic inclusion of both 
$\Xi^- n$ sub-system spin channels. Needless to emphasize that our conclusions tacitly relied on the presumed 
halo-bound structure of the $\Xi^- nn$ system, {\it viz.} a strongly attractive and bound ${}^3S_1$ $\Xi^- n$ 
sub-system and a predominantly weak (attractive or repulsive) ${}^1S_0$ $\Xi^- n$ sub-system with small scattering 
length.  

\vspace{0.1cm}

A more systematic (realistic) approach requires a modification of the dibaryon formalism of the ${}^{{\pi}\!\!\!/}$EFT 
power counting, as well as including subleading order range effects. However, such a modified EFT analysis is beyond 
the scope of the present work and must be certainly explored as a possible future endeavor when more data will become 
available from upcoming experiments, such as ALICE~\cite{Fabbietti:Hyp2018} and FAIR~\cite{SanchezLorente:2014jxa}.    
%For example, a simultaneous perturbative and non-perturbative expansion schemes in the singlet and triplet $\Xi^- n$ sub-system channels respectively, may be incorporated. 
Moreover, a next-to-leading order analysis would involve additional unknown three-body parameters in the theory 
thereby requiring more  empirical inputs which are currently unprocurable.

%%%%%%%%%%%%%%%%%%%%%%%%%%%%%%%%%%%%%%%%%%%%%%%%%%%%%%%%%%%%%%%%%%%%%%%%%%%%%%%%%%%%%%%%%%%%%%%%%%%%%%%%%%%%%%%%%%%%
\begin{acknowledgement}
%%%%%%%%%%%%%%%%%%%%%%%%%%%%%%%%%%%%%%%%%%%%%%%%%%%%%%%%%%%%%%%%%%%%%%%%%%%%%%%%%%%%%%%%%%%%%%%%%%%%%%%%%%%%%%%%%%%%
We wish to acknowledge the support of the Department of Physics, Indian Institute of Technology Guwahati. We thank 
Debades Bandhyopadhyay, Asit Baran Raha and Bipul Bhuyan for various useful comments and discussions. 

\end{acknowledgement}
%

%%%%%%%%%%%%%%%%%%%%%%%%%%%%%%%%%%%%%%%%%%%%%%%%%%%%%%%%%%%%%%%%%%%%%%%%%%%%%%%%%%%%%%%%%%%%%%%%%%%%%%%%%%%%%%%%%%%%
\begin{appendix}
%%%%%%%%%%%%%%%%%%%%%%%%%%%%%%%%%%%%%%%%%%%%%%%%%%%%%%%%%%%%%%%%%%%%%%%%%%%%%%%%%%%%%%%%%%%%%%%%%%%%%%%%%%%%%%%%%%%%
\section{\,\,\,Efimov Physics in low-energy EFT}
%%%%%%%%%%%%%%%%%%%%%%%%%%%%%%%%%%%%%%%%%%%%%%%%%%%%%%%%%%%%%%%%%%%%%%%%%%%%%%%%%%%%%%%%%%%%%%%%%%%%%%%%%%%%%%%%%%%%
For the sake of pedagogical completeness, we highlight aspects of the halo EFT analysis of 
two- and three-body universality responsible for formation of exotic Efimov-like states in {\it fine-tuned} 
three-body systems. {\it Universality} in this context refers to the property of {\it distinct separation of scales}, 
namely, similarities in long-range (low-energy) characteristics of a large class of multi-particle systems which are 
insensitive to the short-distance (high-energy) details. The halo-bound systems typically satisfying this property are
naturally suited for a low-energy EFT description. We elucidate the pertinent EFT framework using the simplest system 
of three {\it identical} interacting spinless bosons ($B-B-B$), where additional involvement of spin and isospin 
degrees of freedom are absent. In particular, we consider a zero-range ``toy-model" (ZRM) scenario (i.e., with two-body
interaction range $r_0\to 0$) which motivates the general formalism of a leading order EFT analysis implemented in 
this paper. The foundations to this framework have been extensively discussed in the context of a large collection of 
pionless EFT (${}^{{\pi}\!\!\!/}$EFT) works in the literature (see e.g.,
\cite{Hammer:2019poc,Braaten:2004rn,Bedaque:1998kg,Bedaque:1999ve,Bedaque:1998km,Kaplan:1996nv,Kaplan:1996xu,Kaplan:1998tg,Kaplan:1998we,Kaplan:1998sz,vanKolck:1998bw,Bedaque:1998mb,Birse:1998dk,Beane:2000fx} and references therein). For the ensuing discussions 
below, we closely follow the reviews works~\cite{Braaten:2004rn,Beane:2000fx,Hjorth-Jensen:2017gss}. 

%%%%%%%%%%%%%%%%%%%%%%%%%%%%%%%%%%%%%%%%%%%%%%%%%%%%%
\subsection{Two-body Sector}
%%%%%%%%%%%%%%%%%%%%%%%%%%%%%%%%%%%%%%%%%%%%%%%%%%%%%
The two-body dynamics stems from the most general non-relativistic effective Lagrangian constituting all possible local short-range 
S-wave two-body interactions of the generic form~\cite{Kaplan:1996nv,Kaplan:1996xu,Kaplan:1998tg,Kaplan:1998we,Kaplan:1998sz}
\begin{equation}
 \mathscr{L}^{\rm (KSW)}_{2} =B^\dagger\bigg[i\partial_t+\frac{\nabla^2}{2m_B}\bigg]B
 -\frac{C_0}{4}\left(B^{\dagger}B\right)^2 
 -\frac{C_2}{4}\left[\nabla(B^{\dagger}B)\right]^2 +\cdots,
 \label{EQ1}
\end{equation}
where  the ellipses denote higher order derivative terms and $C_{0,2,\cdots}$ are two-body coupling constants. These couplings may be 
fixed from the empirical knowledge of the two-body parameters, such as those occurring in the well-known Bethe's {\it Effective 
Range Expansion} (ERE) formula for the S-wave scattering phase-shift $\delta_0$, namely,  
\begin{equation}
k\cot\delta_0 = -\frac{1}{a_0}+\frac{r_0}{2} k^2+\mathcal{O}(k^4)\,, 
\label{EQ4}
\end{equation}  
with $k$ being the center-of-mass scattering momentum of the $B-B$ system. Especially, in the vicinity of shallow two-body (real 
or virtual) bound states these couplings are subject to an {\it unnatural scaling} with the magnitude of the S-wave scattering 
length ($a_0$) becoming unusually large in comparison to the short-distance effective range ($r_0$) of the interactions. In this
case the theory becomes approximately conformally invariant, leading to a peculiar EFT governed by a non-trivial RG fixed point 
of the scale dependent couplings~\cite{Birse:1998dk}.

\vspace{0.1cm}

\begin{figure}[h]
\centering
\resizebox{1.0\columnwidth}{!}{\includegraphics[width=10cm]{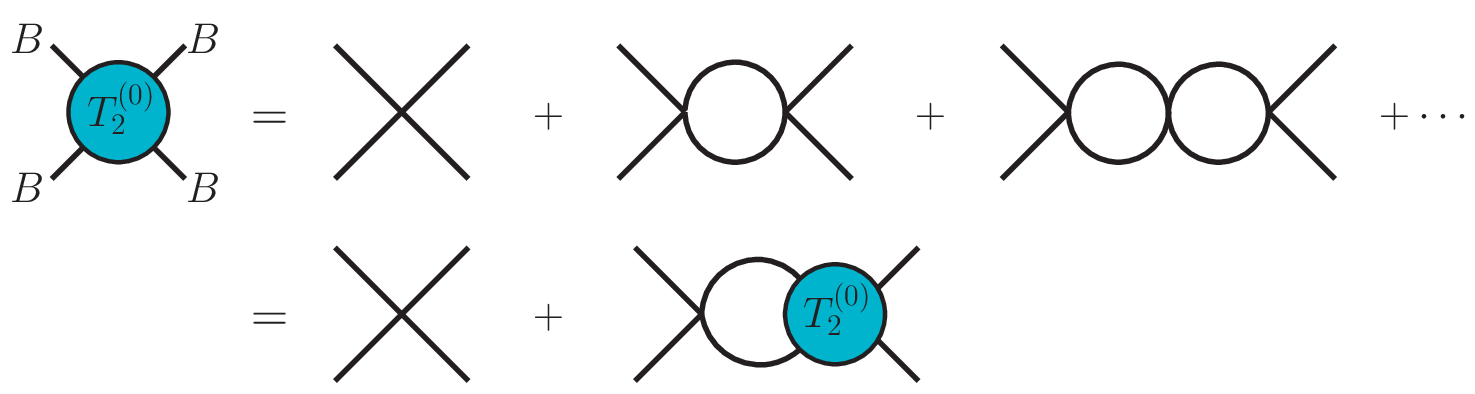}}
    \caption{The leading order two-body S-wave scattering amplitude $T^{(0)}_2$, obtained by resummating the 
             ``bubble" graphs containing the $C_0$ interaction. It is compactly represented as a 
             Lippmann-Schwinger integral equation,
             ${\hat T}^{(0)}_2={\hat V}_2+ {\hat V}_2 {\hat G}_0 {\hat T}^{(0)}_2$.}
\label{fig9} 
\end{figure}

Consider $Q$ to be a generic small momentum associated with a certain emergent low-energy scale $\gamma_0$ of the two-body system, and 
$M_{hi}$ as the UV cut-off or the breakdown scale of the theory associated with the unresolved ``heavy" pion mass $m_\pi$. Then, the 
unnatural scaling scenario implies 
$k \sim Q\sim \gamma_0 \sim 1/|a_0| << M_{hi}\sim 1/r_0$. This is distinguished from the natural scenario where the 
interaction range $r_0\sim 1/m_\pi$ solely accounts for all relevant scales such that 
$k \sim Q\sim \gamma_0 \sim 1/|a_0| \sim M_{hi}\sim 1/r_0$. Such unnatural two-body systems are ubiquitous in nuclear physics, say, in
the case of two neutrons, $a_{nn}= -18.63~{\rm fm}\gg r_{nn}\sim 1/m_\pi\sim 0.5~{\rm fm}$. In all these manifestly {\it fine-tuned} 
scenarios near-threshold (shallow) bound {\it dimer} states emerge by critically tuning the couplings according to  
\begin{equation}
 C_{2r} \sim \frac{1}{2\mu_B M^r_{hi} Q^{r+1}} \quad \forall\, r\in {\mathbb Z}_+ 
\label{EQ7}
\end{equation}
with $\mu_B=m_B/2$ being the reduced mass of the two-boson ($B-B$) system. This kind of two-body contact interactions scaling rule is
termed as the {\it Q-counting} ~\cite{Kaplan:1996xu,Kaplan:1998tg,Kaplan:1998we,Kaplan:1998sz,vanKolck:1998bw}. Based on the
Q-counting different components of Feynman amplitudes scale as follows:
\begin{itemize}
 \item the non-relativistic boson propagator, namely,
 \begin{equation}
     iS_B(p)=\frac{i}{p_0-\frac{{\bf p}^2}{2m_B}+i\eta},
 \end{equation}
 scales as $\sim m_B/Q^2\sim {\mathcal O}(Q^{-2})$,
 \item the loop integral scales as $\sim\mathcal{O} (Q^5)$,~\footnote{This is easily seen as follows:
 \begin{equation*}
\int {\rm d}^4 q \sim \int {\rm d}q_0 \int d^3{\bf q} \sim \frac{Q^2}{2\mu_B}\cdot Q^3 = \frac{Q^5}{2\mu_B}\,.   
\end{equation*}
}
 \item the interaction vertices $C_{2r}\nabla^{2r}$ scale as $\sim\mathcal{O} (Q^{r-1})$.
\end{itemize}
Accordingly, all two-body Feynman graphs contributing up to a fixed order in the counting, say $\mathcal O(Q^N)$, with $L$ loops 
and ${\cal V}_{2r}$ interaction vertices with $2r$ derivatives scale as~\cite{Hjorth-Jensen:2017gss} 
\begin{eqnarray}
T^{(N)}_2 = \sum^N_{\nu = -1} T_2^{(\nu)},\quad T_2^{(\nu)}\sim \mathcal{O} (Q^\nu)\,; 
\quad \nu=3L+2+\sum_{r=0}(r-3){\cal V}_{2r}\geq -1\,.
\label{EQ8}
\end{eqnarray}
Thus, for example at low energies $E\sim 1/a^2_0$, restricting to leading order in Q-counting (i.e., for $N=-1$), an infinite
number of loops with only $C_0$ interactions are needed to be non-perturbatively resummed. Such an infinite sequence of 
``bubble diagrams", all contributing at the same order in the Q-counting (cf. Fig.~\ref{fig9}), yields a shallow two-body bound
state. However, at the next order (i.e., for $N=0$) the $C_2$ interactions contribute to the dynamics. Being $1/M_{hi}$ 
suppressed compared to the leading $C_0$ interactions, the $C_2$ constitutes a perturbative correction. Such a ``bubble 
resummation" with only the $C_0$ interaction leads to a {\it Lippmann-Schwinger series} for the two-body scattering amplitude 
$T^{(0)}_2$ which may be expressed as an integral equation:
\begin{equation}
T^{(0)}_2(E)=-C_0 - \frac{1}{2}C_0 \int \frac{d^4q}{(2\pi)^4 i}\, \frac{i}{q_0-{\bf q}^2/2m_B+i\eta}\,
\frac{i}{E-q_0-{\bf q}^2/2m_B+i\eta}\,T^{(0)}_2(E)\,,
\label{EQ9}
\end{equation}
%%%%%%%%%%%%%%%%%%%%%%%%%%%%%%%%%%%%%%%%%%%%%%%%%%%%%%%%%%%%%%%%%%%%%%%%%%%%%%%%%%%%%%%%%%%%
%%%%%%%%%%%%%%%%%%%%%%%%%%%%%%%%%%%%%%%%%%%%%%%%%%%%%%%%%%%%%%%%%%%%%%%%%%%%%%%%%%%%%%%%%%%%
where $q_0$, ${\bf q}$ are temporal and spatial parts of loop four-momentum $q$. The integral equation may be regularized 
using a sharp momentum cut-off $\Lambda_{\rm reg}$ and subsequently solving to obtain
\begin{equation}
T^{(0)}_2 (E,\Lambda_{\rm reg}) = -C_0 \left[1+\frac{\mu_B C_0}{2\pi^2}\left(\Lambda_{\rm reg}
-\frac{\pi}{2}\sqrt{-2\mu_B E-i\eta}\right)\right]^{-1}.
\label{EQ10}
\end{equation}
The renormalization of the amplitude $T^{(0)}_2(E)$ can be implemented by matching to the two-body S-wave scattering 
length using the relation 
\begin{equation}
a_0 =-\frac{\mu_{B}}{4\pi}\lim_{k\rightarrow 0} T^{(0)}_2(E={\bf k}^2/2\mu_B)\,,
\end{equation}
leading to the regulator dependent result for the coupling $C_0$, namely,
\begin{eqnarray}
C_0(\Lambda_{\rm reg}) = \frac{4\pi a_0}{\mu_B} \left(1-\frac{2a_0\Lambda_{\rm reg}}{\pi}\right)^{-1}\,,
\label{EQ11}
\end{eqnarray}
such that the renormalized scattering amplitude is given by
\begin{eqnarray}
T_2 (E) = \frac{4\pi}{\mu_B} \frac{1}{-1/a_0+\sqrt{-2\mu_B E-i\eta}}\,.
\end{eqnarray}

{\it Auxiliary Field Formalism:} In the context of studying three-body dynamics with two-body bound subsystems with large scattering lengths, it is convenient to introduce auxiliary {\it dimer} fields or 
{\it dimerons} ($d$). Ideally, the bare auxiliary fields do not propagate and even have a ``wrong sign" in the 
kinetic terms (see following two-body Lagrangian). Hence, the content of the original theory remains unchanged by the addition
of such ghost fields ``by hand".  For the present case of the spinless three-boson system ($B-B-B$), an alternative 
two-body Lagrangian in terms of the dimeron fields can be 
constructed as~\cite{Bedaque:1998kg,Bedaque:1999ve,Bedaque:1998km},  
\begin{eqnarray}
 \mathscr{L}^{\rm (BHvK)}_2 \!=\! B^{\dagger}\bigg[i\partial_t+\frac{\nabla^2}{2m_B}\bigg]B
-d^{\dagger}\bigg[i\partial_t+\frac{\nabla^2}{4m_B}-\Delta_d\bigg]d 
-y_0 \! \left(d^{\dagger}B^2+B^{\dagger 2} d\right) +\cdots.\,\quad\,
 \label{EQ13}
\end{eqnarray}
In effect, the dimerons are essentially employed to cancel the quadratic terms such as $(B^{\dagger}B)^2$ in 
Eq.~\eqref{EQ1}, so that all interactions between the $B$ fields are now mediated {\it via} the dimer exchange 
process, with $y_0$ as the corresponding interaction coupling. The quantity $\Delta_d$ is a free parameter related to
the binding energy of the dimeron $d$ such that the bare or tree-level dimeron propagator is the simple non-dynamical
term $i/\Delta_d$. Quantum loop corrections, however, allow the dimerons to propagator. It is notable that by virtue 
of reparametrization invariance of the theory, the above Lagrangian can be shown to be equivalent to the original 
two-body Lagrangian, Eq.~\eqref{EQ1}.

\vspace{0.1cm}

In the context of halo EFT where the dimeron formalism has been extensively used, the Q-counting scheme has been extended 
to include the scaling,
\begin{equation}
y^2_0\sim \frac{M_{hi}}{4\mu^2_B}\sim {\mathcal O}(1) \qquad \text{and} 
\qquad \Delta_d\sim \frac{M_{hi} Q}{2\mu_B}\sim {\mathcal O}(Q)\,.
\end{equation}
Consequently, using field re-definitions in trading away the time derivatives in favor of space derivatives, the kinetic 
term becomes sub-leading compared to term proportional to $\Delta_d$. In that case the Q-counting leads to an infinite 
sequence of Feynman graphs (similar to the ones displayed in Fig.~\ref{fig-7}), each contributing at the same order as the
static dimeron propagator $i\Delta^{-1}_d \sim {\mathcal O}(Q^{-1})$. Consequently, they must all be resummed at the 
leading order to yield the full dynamical ``dressed" dimeron propagator:
\begin{equation}
 i\Delta^{(0)}(p_0,{\bf p})= \frac{i\pi}{y^2_0 \mu_B} 
 \left[\frac{\pi \Delta_d}{y^2_0 \mu_B}+\frac{2}{\pi}\Lambda_{\rm reg}-\sqrt{-2\mu_B p_0+{\bf p}^2/4-i\eta}\right]^{-1}.
 \label{EQ15}
\end{equation}
We note its similarity to the resummed two-body scattering amplitude, Eq.~\eqref{EQ10}, with the two-body center-of-mass 
kinetic energy $E\to p_0-{\bf p}^2/(8\mu_B)$, where $p_0 ({\bf p})$ is kinetic energy (momentum) of the dimeron. Upon 
renormalization using Eq.~\eqref{EQ11} and the leading order relation among the two-body parameters, namely, 
$C_0 \to 4y^2_0/\Delta_d$, the renormalized dressed dimeron propagator becomes
\begin{equation}
 i\Delta(p_0,{\bf p}) = -\frac{i\pi}{y^2_0 \mu_B} \left[-\frac{1}{a_0}+\sqrt{-2\mu_B p_0+{\bf p}^2/4-i\eta}\right]^{-1}.
 \label{EQ16}
\end{equation}
If scattering length $a_0>0$, then the above dimeron propagator has a pole at $p_0 = -1/(2\mu_B a^2_0) + {\bf p}^2/(8\mu_B)$. 
For low-energy threshold processes, ${\bf p}\rightarrow 0$ and $p_0\rightarrow -{\mathcal B}_2=-1/(2\mu_B a^2_0)$, in which 
case the renormalized dimeron propagator has a residue (wave function renormalization constant) at the pole given by
\begin{eqnarray}
 {\mathcal Z}_d = \left[\frac{{\rm d}\Delta^{-1}_d(p_0,{\bf 0})}{{\rm d}p_0}\right]^{-1}_{p_0=-{\mathcal B}_2}
 =\frac{2\pi}{y^2_0\mu^2_B a_0}\,.
 \label{EQ16A}
\end{eqnarray}

%%%%%%%%%%%%%%%%%%%%%%%%%%%%%%%%%%%%%%%%%%%%%%%%%%%%%
\subsection{Three-body Sector}
%%%%%%%%%%%%%%%%%%%%%%%%%%%%%%%%%%%%%%%%%%%%%%%%%%%%%
%%%%%%%%%%%%%%%%%%%%%%%%%%%%%%%%%%%%%%%%%%%%%%%%%%%%%%%%%%%%%%%%%%%%%%%%%%%%%%%%%%%%%%%%%%%%
\begin{figure}[bp]
\centering
\resizebox{1.0\columnwidth}{!}{\includegraphics{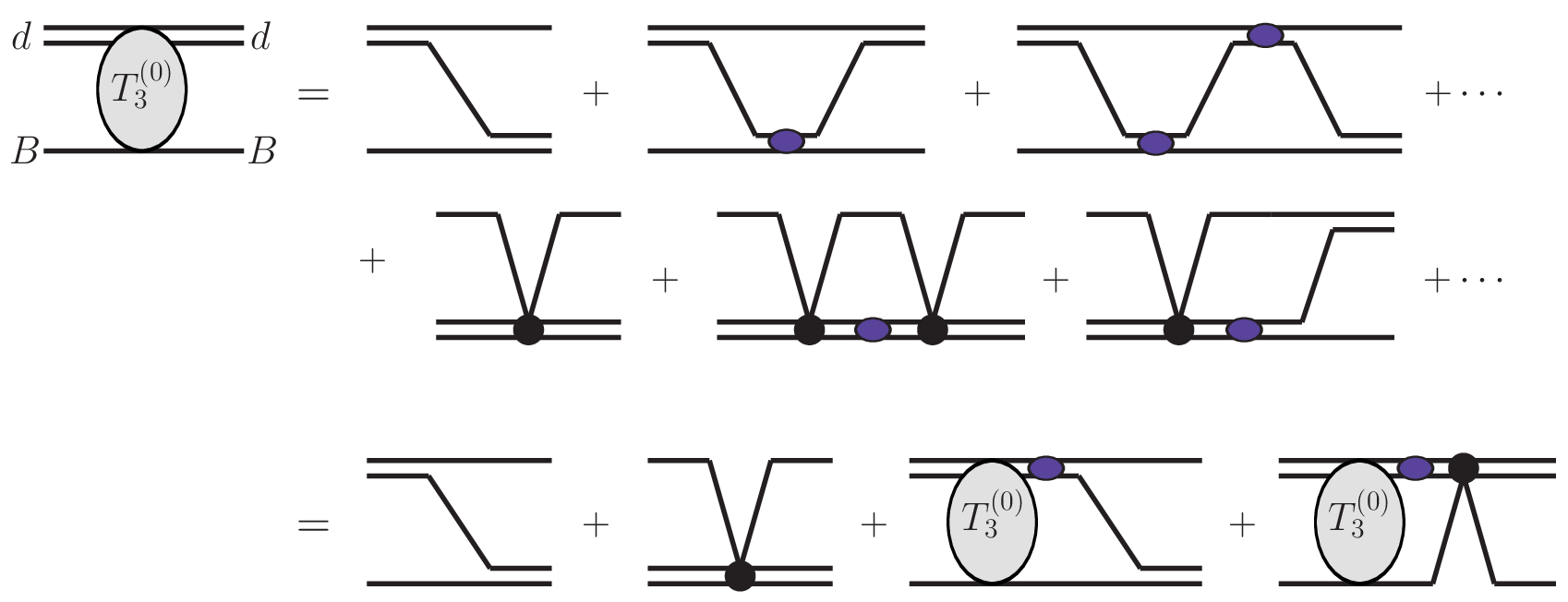}}
    \caption{Diagrammatic representation of the three-body integral equation for the spinless three-boson S-wave scattering amplitude 
             $T^{(0)}_3$. In the Q-counting scheme, all graphs in the first line contribute as $\sim M_{hi}/(2\mu_B Q^2)$, 
             while those in the second line with three-body contact interactions contribute as $\sim 1/(2\mu_B Q^4)$. The 
             single line denotes a boson ($B$) propagator, the double line denotes a static dimeron ($d$) propagator, and the
             double line with an oval blob represents a fully dressed (dynamical) renormalized dimeron propagator. Finally, 
             the dark filled circle represents an insertion of a leading order three-body contact interaction. }
\label{fig11} 
\end{figure}
%%%%%%%%%%%%%%%%%%%%%%%%%%%%%%%%%%%%%%%%%%%%%%%%%%%%%%%%%%%%%%%%%%%%%%%%%%%%%%%%%%%%%%%%%%%%
Here we describe the dynamics associated with S-wave scattering of the three-boson system depicted in Fig.~\ref{fig11}. 
Excluding the {\it genuine} three-body interaction (3BF) diagrams shown in the second row, the scattering diagrams in 
the first row constitute only the two-body interaction $y_0$, and thus described by same leading order EFT Lagrangian, 
Eq.~\eqref{EQ13}. Under the Q-counting scheme, a simple investigation shows that these diagrams contribute at the same order. 
For example, the tree level diagram in first row can be considered as a one-boson ($B$) exchange process between the incoming 
and outgoing dimerons ($d$). Since the tree diagram containing two $dBB$ vertices ($y^2_0\sim M_{hi}/4\mu_B^2$) and one $B$ 
propagator ($iS_B\sim m_B/Q^2$), this amounts to a net contribution of order $\sim M_{hi}/(2\mu_B Q^2)$. The adjacent one-loop 
re-scattering diagram contains two additional $dBB$ vertices, two additional $B$ propagators, a dimeron propagator 
($i\Delta \sim 2\mu_B/M_{hi} Q$), and a loop integral ($\sim Q^5/2\mu_B$). This implies that the one-loop diagram also scales 
as $\sim M_{hi}/(2\mu_B Q^2)$, like the tree graph. In fact, inspection shows that every re-scattering graph in the first row 
precisely contributes at the same leading order, necessitating a resummation of all such graphs. Besides, one also requires
the infinite sequence of 3BF diagrams (second row) which may be shown to contribute at the leading order, and hence needed to 
be resummed as well. The power counting for such graphs is described later in this section. However, unlike the resummation 
in the two-body sector [cf. Eq.~\eqref{EQ9}] which amounts to a simple summation of a geometric series, the same is 
impracticable in the three-body sector. In contrast, here we must incorporate a non-perturbative resummation in the form of a 
``one-loop" {\it Fredholm} integral equation (third row of Fig.~\ref{fig11}) which is solved self-consistently for the 
scattering amplitude $T^{(0)}_3$ using numerical methods. It is notable that the standard 
{\it Faddeev equations}\,\footnote{The Faddeev equations, as distinct from the three-particle Schr{\"o}dinger equation, is a 
set of three coupled channel equations tailor-made to exploit configurations consisting of a two-body cluster that is 
well-separated from the third particle, leading to considerable simplifications in the solution. Moreover, at low-energies 
ignoring the subsystem angular momentum follows more naturally than in Schr{\"o}diger equation. With considerable 
simplifications they reduce to a single integro-differential equation for the three-boson wavefunction $\psi(R,\alpha)$, 
with {\it hyperradius} $R$ and generic {\it Delves' hyperangle} $\alpha \equiv \alpha_k$ , defined by
$R=\frac{1}{2}r^2_{ij}+\frac{2}{3}r^2_{k,ij}$ and $\alpha_k={\rm arctan}\left(\sqrt{3}r_{ij}/2r_{k,ij}\right)$ respectively, in 
Jacobi coordinates, where $(i,j,k)$ is a cyclic permutation of $(1,2,3)$. This leads to the so-called {\it low-energy Faddeev 
equation}~\cite{Braaten:2004rn,Fedorov93}:
\begin{eqnarray}
    (T_R+T_\alpha -E)\psi(R,\alpha)=-V(\sqrt{2}R\sin\alpha)\!\bigg[\psi(R,\alpha)&+&\frac{4}{\sqrt{3}}\!
    \int^{\frac{\pi}{2}-|\frac{\pi}{6}-\alpha|}_{|\frac{\pi}{3}-\alpha|}\!
    \frac{\sin(2\alpha^\prime)}{\sin(2\alpha)}\psi(R,\alpha^\prime) {\rm d}\alpha^\prime\bigg]\,,
    \nonumber \\
    \text{with}\quad T_R=-\frac{1}{4\mu_B}\left[\frac{\partial^2}{\partial R^2}+\frac{5}{R}\frac{\partial}{\partial R}\right]\,\,,\quad 
    T_\alpha&=&-\frac{1}{4\mu_B R^2}\left[\frac{\partial^2}{\partial \alpha^2}+4\cot(2\alpha)\frac{\partial}{\partial \alpha}\right]\,.
    \nonumber
\end{eqnarray}
} are a set of coupled {\it integro-differential} equations in the co-ordinate representation. They are numerically solved using
finite range model potentials with well-defined kernels in the Hilbert-Schmidt class. Our ZRM integral equations in contrast have 
an ill-defined kernel stemming from the non-self-adjoint character of the underlying three-body 
Hamiltonian~\cite{Danilov:1961,Faddeev:1961bg}. Such ``Faddeev-like" zero range three-body integral equations in the momentum 
representation are termed as the STM or the so-called {\it Skornyakov-Ter-Martirosyan} equations~\cite{STM1,STM2}. Despite the 
ambiguities in their solution, the STM equations have the primary advantage of fitting naturally into an EFT framework based 
on a diagrammatic or Lagrangian-based approach. This contrasts with the inherently non-perturbative Hamiltonian-based potential 
model approach for solving three-body Faddeev equations. 

\vspace{0.1cm}

The biggest advantage, however, lies in the manner of introducing 3BF  which is quite naturally achieved in the STM 
framework without the requirement of {\it ad hoc} three-body potentials. The ambiguity in the Hamiltonian is solved by 
cutting off the effective interactions at short-distances. This is accomplished in the EFT by introducing, e.g., a 
sharp momentum cut-off $\Lambda_{\rm reg}$ in the integral equations which becomes a free parameter of the theory. This 
regularization method implicitly necessitates the introduction of cut-off dependent three-body couplings needed to 
renormalize the artificial cut-off dependence of the STM equations. To this end, we introduce additional 
non-derivatively coupled three-body interactions in the effective Lagrangian, for instance,
\begin{eqnarray}
\mathscr{L}^{\rm (KSW)}_{3} &=&  \frac{D_0(\Lambda_{\rm reg})}{36} (B^\dagger B)^3 +\cdots\,, 
\qquad \text{(standard form)} \nonumber
\end{eqnarray}
\begin{eqnarray}
\mathscr{L}^{\rm (BHvK)}_{3} &=&  -\,d_0(\Lambda_{\rm reg}) \left(d^\dagger d\right)\left(B^\dagger B\right) +\cdots\,, 
\qquad \text{(dimerized form)} 
\label{EQAPPL3}
\end{eqnarray}
with the contact interaction couplings scaling unnaturally at the leading order, namely, $D_0\sim d_0 \sim 1/(2\mu_B Q^4)$. 
In this case the $B-B-B$ system exhibits three-body (Efimov) universality.\footnote{For {\it natural} systems without three-body 
universality, the scaling of the 3BF couplings are instead governed by {\it naive} dimensional analysis (NDA), 
$D_0\sim d_0 \sim 1/(2\mu_B M^4_{hi})$. Thus, the corresponding 3BF terms are considered subleading in the EFT Lagrangian.} 
The values of the 3BF couplings can be fixed using additional three-body datum (e.g., three-body binding energy) in realistic 
situations. The ellipses denote derivative 3BF terms which are naturally subleading. Figure~\ref{fig11} (second row) displays 
all re-scattering diagrams with insertions of leading order three-body contact interactions. These graphs are similarly 
resummed into an integral equation that yields contributions of order $\sim 1/(2\mu_B Q^4)$, and hence equally important 
as the set of graphs in the first row with two-body interactions only. Together they constitute all possible non-perturbative 
contributions to the scattering amplitude at leading order.    

\vspace{0.1cm}

Assuming the manifestation of Efimov universality, the natural choice of the reference frame is the boson-dimeron ($B-d$)
center-of-mass (CM), with the relative external momenta being ${\bf -k,k}$ (${\bf -p,p}$) for the incoming (outgoing) boson 
and dimeron respectively. With the total three-body CM kinetic energy as $E$, their energies are taken as $E_A,\, E-E_A$ 
($E'_A,\, E-E'_A$) for the incoming (outgoing) particles. Using standard Feynman rules which follow from the EFT Lagrangians,
Eqs.~\eqref{EQ13} and \eqref{EQAPPL3}, we easily obtain the following Faddeev-like STM integral equation for the unrenormalized 
scattering amplitude $T^{(0)}_3$ corresponding to the second line of Fig.~\ref{fig11}, namely,
\begin{eqnarray}
 T^{(0)}_3\left({\bf p,k};E\right) &=& -\left[\frac{4y^2_0}{E-E'_A-E_A-({\bf p+k})^2/2m_B +i\eta}+d_0(\Lambda_{\rm reg})\right] 
 \nonumber\\
 && +\, \frac{i\pi}{\mu_By^2_0}\int^{\Lambda_{\rm reg}} \frac{{\rm d}^4q}{(2\pi)^4} 
 \bigg[\frac{4y^2_0}{E-E'_A-q_0- ({\bf p+q})^2/2m_B+i\eta }+ d_0(\Lambda_{\rm reg})\bigg] 
 \nonumber\\
 && \hspace{1cm} \times\, \frac{1}{q_0-{\bf q}^2/2m_B +i\eta}\,
 \frac{T^{(0)}_3\left({\bf q,k};E\right)}{1/a_0-\sqrt{-2\mu_B(E-q_0)+{\bf q}^2/4-i\eta}},
 \label{EQ17}
\end{eqnarray}
where $q_0 ({\bf q})$ is the temporal (spatial) part of loop momentum with the three-momentum integral cut-off in the UV 
region at $|{\bf q}| = \Lambda_{\rm reg}$. Using Cauchy's residue theorem, the integral over $q_0$ can be evaluated by 
choosing the pole, $q_0={\bf q}^2/2m_B$, making one of the boson propagators inside the loop integration on-shell. Further
simplifications is achieved by choosing either one or both the initial and final states on-shell. For instance, with
the full on-shell choice $E_A = {\bf k}^2/2m_B$ and $E'_A= {\bf p}^2/2m_B$, we obtain 
\begin{eqnarray}
 T^{(0)}_3\left({\bf p,k};E\right) &=& -4m_By^2_0 
 \left[\frac{1}{m_B E-({\bf p}^2+{\bf p \cdot k}+{\bf k}^2)+i\eta} +\frac{d_0(\Lambda_{\rm reg})}{4m_By^2_0}\right] 
 \nonumber \\
 && -\, 8\pi \int^{\Lambda_{\rm reg}} \frac{{\rm d}^3{\bf q}}{(2\pi)^3} 
 \bigg[\frac{1}{m_B E-({\bf p}^2+{\bf p \cdot q}+{\bf q}^2)+i\eta} 
 +\frac{d_0(\Lambda_{\rm reg})}{4m_By^2_0}\bigg] 
 \nonumber\\
 && \hspace{2.5cm} \times\, \frac{ T^{(0)}_3\left({\bf q,k};E\right) }{-1/a_0+\sqrt{-2\mu_B E+3{\bf q}^2/4-i\eta}}\, .
 \label{EQ18}
\end{eqnarray}
Especially in the context of S-wave scattering process, we consider the projection of the unrenormalized amplitude onto
the $l=0$ partial wave renormalized amplitude given by
\begin{equation}
T_3\left(p,k;E\right) \equiv \frac{{\mathcal Z}_d}{2} \int_{-1}^1 {\rm d}
\left(\cos\theta_{\hat{\bf p}\cdot \hat{\bf q}}\right)\,\,T^{(0)}_3\left({\bf p,k};E\right)\,,
\label{EQ19}
\end{equation}
such that boson-dimeron {\it elastic} scattering amplitude is obtained by evaluating the renormalized scattering 
amplitude at the on-shell point, $p=|{\bf p}|=k=|{\bf k}|$ and $E=-{\mathcal B}_2+3k^2/(4m_B)$, with dimer binding 
energy ${\mathcal B}_2=1/(2\mu_Ba^2_0)$. Furthermore, in the threshold limit one obtains the three-body scattering length
as
\begin{eqnarray}
 a^{(Bd)}_3= -\frac{m_B}{3\pi}\lim_{k\to 0}\,\, T^{(0)}_3\bigg(k,k;\frac{3k^2}{4m_B}-{\mathcal B}_2\bigg)\,.
\end{eqnarray}
It is conventional to re-define the dimension-full three-body coupling $d_0(\Lambda_{\rm reg})$ in terms of a 
dimensionless coupling $H(\Lambda_{\rm reg})$ such that $T^{(0)}_3(p,k;E)$  has a well-defined asymptotic behavior as 
$\Lambda_{\rm reg} \rightarrow \infty$, namely,
\begin{equation}
 d_0(\Lambda_{\rm reg}) = -\frac{4m_By^2_0}{\Lambda^2_{\rm reg}} H(\Lambda_{\rm reg})\,.
\label{EQ20}
\end{equation}
This leads to the 3BF renormalized STM integral equation for the three-boson system, originally derived in 
Ref.~\cite{Bedaque:1998km}:
\begin{eqnarray}
 T_3(p,k;E) &=& \frac{8\pi}{\mu_B a_0}\left[ \frac{1}{2pk}\ln \left(\frac{p^2+k^2+pk-m_B E-i\eta}
 {p^2+k^2-pk-m_B E-i\eta}\right)+\frac{H(\Lambda_{\rm reg})}{\Lambda^2}\right]
 \nonumber\\
 &&+\,\frac{4}{\pi}\int_0^{\Lambda_{\rm reg}} dq\,\, q^2\left[ \frac{1}{2pq}\ln \left(\frac{p^2+q^2+pq-m_B E-i\eta}
 {p^2+q^2-pq-m_B E-i\eta}\right)+\frac{H(\Lambda_{\rm reg})}{\Lambda^2}\right]
 \nonumber\\
 && \hspace{2.5cm} \times\, \frac{T_3(q,k;E)}{-1/a_0+\sqrt{3q^2/4-\mu_B E-i\eta}}.
 \label{EQ21}
\end{eqnarray}
The above equation must be numerically solved to obtain the three-body eigenenergies and scattering lengths in the 
respective kinematical domains.\footnote{It must be understood that the three-body bound states are obtained in the 
negative energy kinematical region, $E< {\mathcal E}_{d}$, namely below the boson-dimeron breakup threshold 
${\mathcal E}_d\sim -{\mathcal B}_2$. While, the $B-d$ scattering solutions correspond to energies, 
${\mathcal E}_{d}\leq E< 0$, namely the kinematical region in between the boson-dimeron and three-boson breakup 
thresholds.} In this case the STM equation is one-dimensional in the sense that $B-d$ scattering involves a single
channel elastic process $B+d\to B+d$. In realistic situations with non-zero spin-isospin degrees of freedom, the 
processes involve coupled elastic and inelastic channels, thereby requiring multi-dimensional representations. For 
instance, the $\Xi^- nn$ ($I=3/2,\, J^P={1/2}^+$) system dealt in this paper involves a system of three 
coupled-channel scattering processes: the elastic channel $n+(\Xi^-n)_t\to n+(\Xi^-n)_t$, and the two inelastic 
channels $n+(\Xi^-n)_t\to n+(\Xi^-n)_s$ and $n+(\Xi^-n)_t\to \Xi^-+(nn)_s$, where the subscripts represent the 
sub-system spins. However, for the sake of simplicity the former inelastic channel involving the spin-singlet dimer
$(\Xi^-n)_s$ is assumed to be decoupled (see text). Consequently, we deal with a reduced system of two 
coupled-channel integral equations in terms of the {\it half-on-shell} amplitudes $t^{(R)}_{A,B}$ (cf. Appendix B).
Moreover, additional re-coupling coefficients are necessary for projecting each of the scattering diagrams onto the 
correct spin-isospin channels.

%%%%%%%%%%%%%%%%%%%%%%%%%%%%%%%%%%%%%%%%%%%%%%%%%%%%%
\subsection{RG Limit Cycle}
%%%%%%%%%%%%%%%%%%%%%%%%%%%%%%%%%%%%%%%%%%%%%%%%%%%%%
With sufficiently large $B-B$ scattering length, i.e., $a_0\to \infty$, and very short range two-body interactions, 
$r_0\to 0$, three-body or Efimov universality implies the existence of a tower of arbitrarily-shallow three-body 
bound states close to the unitary or resonant limit as $\Lambda_{\rm reg}\to \infty$. This remarkable discovery is 
credited to Vitaly Efimov~\cite{Efimov:1970zz,Efimov:1971zz}, who in 1970 demonstrated that the system of three
identical bosons interacting {\it via} attractive (i.e, with $a_0>0$) inverse square channel potential 
$V \sim -\left(s^2_0+1/4\right)/(4\mu_BR^2)$ becomes resonant. By solving the Faddeev equations using the well-known 
{\it hyperspherical} representation in coordinate space with suitable short-distance {\it adiabatic} boundary 
conditions, a geometric sequence of three-particle level states (Efimov states) was obtained. The corresponding binding energies 
$B^{(n)}_3$ were found to be lie approximately within the interval,
\begin{eqnarray}
 \frac{1}{2\mu_B a^2_0} \lesssim B^{(n)}_3 \stackrel{n\to \infty}{\longrightarrow }
 \frac{\kappa^2_*}{2\mu_B} \left(e^{-2\pi/s_0}\right)^{(n-n_*)} \lesssim \frac{1}{2\mu_B r^2_0}\,,
\end{eqnarray}
where $n=n_*$ is some integer labeling for a reference level with binding momentum $\kappa_*$, determined by the 
{\it Efimov spectrum} in the (asymptotic) unitary limit. The multiplicative factor $\lambda_0 \equiv e^{\pi/s_0}$
is an {\it universal parameter} which depends only on the gross features of the three-body system, such as the mass 
ratios of the bound particles and the overall quantum statistics of the system, irrespective of the fine details such
as the nature of the individual bound particles and the short-distance interaction potentials. In this case, 
$s_0=1.00624\cdots$ is obtained as a solution to the transcendental equation 
\begin{equation}
s_0\cosh\frac{\pi s_0}{2}=\frac{8}{\sqrt{3}}\sinh\frac{\pi s_0}{6}\,.
\label{EQAPPs0}
\end{equation}
It is notable that mass ratios play the most crucial role in deciding the typical estimate for $s_0$. However, the 
spin, isospin and other possible internal quantum numbers of the gross three-body system can fine-tune its precise 
value. Moreover, away from unitarity in the non-resonant domain as $a_0\sim r_0$, the value of $s_0$ is likely to change 
from its asymptotic value due to cut-off and other low-energy parametric dependencies thereby becoming non-universal.
For instance, the three-nucleon ($N-N-N$) iso-doublet S-wave system of {\it triton} and {\it helion} (${}^{3}$H, 
${}^{3}$He) in the $I=J=1/2$ channel, are probably the best known examples of realistic Efimov-bound states in nuclear 
physics having identical asymptotic scale parameter $s_0$ and Efimov spectrum (neglecting Coulomb interactions) as in 
the case of the $B-B-B$ system.   

\vspace{0.1cm}

\begin{figure}[tbp]
\centering
\includegraphics[width=6.3cm]{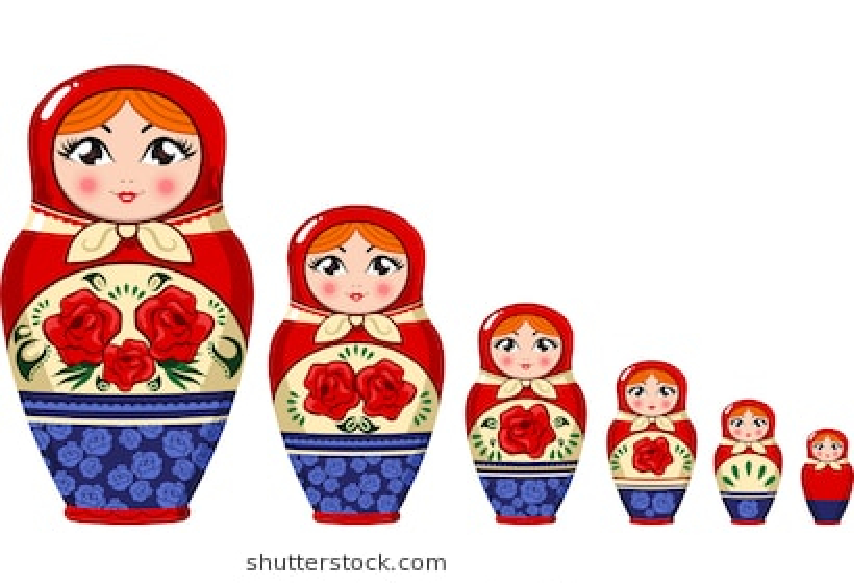}~\includegraphics[width=6.5cm]{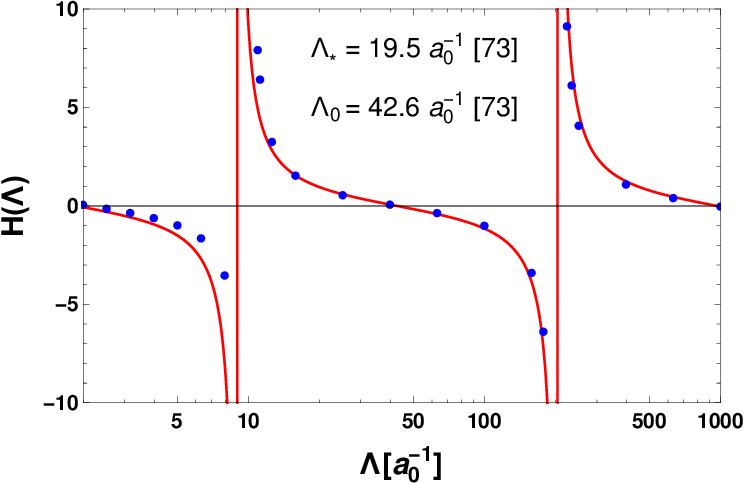}
    \caption{Demonstration of RG limit cycle. {\bf Left panel:} Discrete scaling behavior found in Russian nesting dolls  
             with sizes of successive dolls decreasing by a constant factor $\sim 1.5$. {\bf Right panel:} The regulator 
             scale $\Lambda_{\rm reg}$ dependence of the three-body coupling $H(\Lambda_{\rm reg})$ for the $B-B-B$ system.
             The input three-body datum is the scattering length $a^{(Bd)}_3=1.56 a_0$. The parameters, $\Lambda_*$ and 
             $\Lambda_0$, are obtained by fitting Eq.~\eqref{EQ23} (solid curve) to the data points obtained by numerically
             solving the STM equation~\eqref{EQ21}, reproducing the result of Ref.~\cite{Bedaque:1998km}.}
\label{fig13} 
\end{figure}

The above-mentioned behavior of the three-body system in the so-called {\it scaling limit} ($r_0\to 0$) is an indication 
that the familiar {\it continuous scaling} symmetry (conformal invariance)  under the scale transformation
\begin{eqnarray}
a_0 \rightarrow \lambda a_0 \qquad \text{and} \qquad E_{\rm 2-body}\rightarrow \lambda^{-2} E_{\rm 2-body},
\label{EQ22}
\end{eqnarray}
where $\lambda\in {\mathbb R}_+$ is an arbitrary constant, gets evidently broken in the three-body sector into the 
{\it discrete scaling} subgroup of scale transformations given by
\begin{equation}
\kappa_* \rightarrow \kappa_*\, \,\,\, a_0\rightarrow \lambda^{n}_0 a_0\,, \,\,\,
E\rightarrow \lambda^{-2n}_0 E,\,\,\, \text{where}\,\,\, \lambda_0=e^{\pi/s_0}\approx 22.7\,.
\end{equation}
This feature can be attributed to the introduction of the parameter $\kappa_*$ which sets a new relevant scale in 
the three-body system, in addition to the only existing relevant scale set by the two-body parameter $a_0\to \infty$ 
close to the unitary limit. This leads to logarithmic scaling violation of low-energy observables which must scale as
some log-periodic function $\sim f\left[s_0\ln(\kappa_*|a_0|)\right]$ . In other words, the expected non-trivial RG 
fixed point scaling of the 3BF couplings breaks down into that of an UV RG limit cycle discrete scaling characterized 
by the parameter $\lambda_0$. This unusual type of RG arises in other branches of physics as well, such
as in condensed matter (see e.g., Refs.~\cite{Glazek:2002hq,Glazek:2004zz,LeClair:2002ux,Leclair:2003xj,LeClair:2003hj}),
or in the study of turbulence and complex systems (see e.g., Ref.~\cite{Sornette:1997pb}). Curiously enough, such a discrete 
scaling behavior bears close resemblance to the well-known Russian {\it Matryoshka} dolls, as displayed in 
Fig.~\ref{fig13}. They consist of an assembly of hollow wooden dolls of decreasing size nested one within the other such that the 
ratio of the sizes of successive dolls remains approximately constant, for instance, as shown in the figure, the discrete
scaling factor is given by  
\begin{equation}
e^{2\pi/s^{\rm (doll)}_0} \approx \frac{{\rm doll}^{(n)}}{{\rm doll}^{(n+1)}} \approx   1.5\,.
\end{equation}
Likewise, in proximity to the unitary limit, the ratio of the successive binding energies of the $B-B-B$ Efimov {\it trimer} levels 
scale as $B^{(n)}_3/B^{(n+1)}_3 \approx e^{2\pi/s_0} \approx 515$. The approximate relation 
becomes an exact one only in the unitary limit with $a_0=\infty$.

\vspace{0.1cm}

The RG limit cycle features associated with Efimov spectrum are deduced quite naturally in low-energy EFT by studying 
the cut-off regulator dependence of the 3BF coupling $H(\Lambda_{\rm reg})$ {\it via} the STM integral equation~\eqref{EQ21}. 
The introduction of UV regulator $\Lambda_{\rm reg}$ repairs the non-self-adjoint pathology associated with the STM 
equation making the scattering amplitude $T_3$ well-behaved asymptotically. But this comes at a cost: the continuous scaling symmetry gets 
partially broken, leading to the emergence of an RG limit cycle. This feature can be checked analytically by 
investigating the asymptotic nature of the integral equation in the unitary and scaling limits. In other words, with $p$ 
taken as the off-shell (outgoing) momentum having the same order of magnitude as the loop momenta $q$, we examine the 
integral equations in the limit $E,1/|a_0|,k \ll p\sim q \sim \Lambda_{\rm reg} \lesssim \infty$, whereby the 3BF terms 
$\propto H(\Lambda_{\rm reg})/\Lambda^2_{\rm reg}$ may be dropped. To that effect, the asymptotic solution for $T_3$ scales
as a pure power-law with an undetermined exponent, $T_3 (p)\sim p^{s-1}$. Thus, 
the resulting STM equation~\eqref{EQ21} becomes
\begin{equation}
p^{s-1} = \frac{4}{\sqrt{3}\pi p}\int_0^{\infty} {\rm d}q \,\,q^{s-1}\ln\frac{p^2+pq+q^2}{p^2-pq+q^2}\,,
\label{EQ24}
\end{equation}
which after a change of variable, $q=xp$, becomes
\begin{equation}
1 = \frac{4}{\sqrt{3}\pi} \int_0^{\infty} dx \,\, x^{s-1} \ln \frac{1+x+x^2}{1-x+x^2}\,.
\label{EQ25}
\end{equation}
A Mellin transformation finally reduces the STM equation to the very aforementioned transcendental relation, Eq.~\eqref{EQAPPs0},
obtained by solving the low-energy Faddeev equation, albeit with a complex exponent $s=\pm is_0$, where $s_0=1.00624\cdots$
is obtained in this case. The asymptotic value $s_0$ parametrizes the exact {\it discrete scaling} behavior at the unitary limit, 
formally indicating the manifestation of Efimov effect. Furthermore, Bedaque {\it et al.}~\cite{Bedaque:1998km} deduced an approximate 
analytical expression for the typical log-periodic running of dimensionless three-body coupling $H(\Lambda_{\rm reg})$, 
given by
\begin{eqnarray}
 H(\Lambda_{\rm reg}) = -\frac{\sin\left[s_0 \ln(\Lambda_{\rm reg}/\Lambda_*)
 -\arctan(1/s_0) \right]}{\sin\left[s_0 \ln(\Lambda_{\rm reg}/\Lambda_*)+\arctan(1/s_0) \right]}\,.
 \label{EQ23}
\end{eqnarray}
%%%%%%%%%%%%%%%%%%%%%%%%%%%%%%%%%%%%%%%%%%%%%%%%%%%%%%%%%%%%%%%%%%%%%%%%%%%%%%%%%%%%%%%%%%%%
%%%%%%%%%%%%%%%%%%%%%%%%%%%%%%%%%%%%%%%%%%%%%%%%%%%%%%%%%%%%%%%%%%%%%%%%%%%%%%%%%%%%%%%%%%%%
Such an RG orbit for the coupling constant with a periodic dependence on the cut-off parameter when the latter 
increases to infinity is termed as a limit cycle. The underlying principle ensures that the family of 
effective theories with finite cut-offs yields predictions which are guaranteed to remain independent of the respective 
cut-offs. In the above expression $\Lambda_*$, in analogy to $\kappa_*$, is an emergent three-body 
dynamical parameter which results from the logarithmic scaling violation $\sim \ln(\Lambda_*|a_0|)$. This 
parameter is likewise fixed using a three-body datum, such as the trimer binding energy $B_3$ or the $B-d$ 
scattering length $a^{(Bd)}_{3}$. Alternatively, the scale dependence of $H(\Lambda_{\rm reg})$ may be 
determined by numerically solving the STM equation~\eqref{EQ21} at non-asymptotic scales for a given two-body 
input $a_0$ and three-body datum (e.g., $B_3,\,a^{(Bd)}_3$, etc.). In Fig.~\ref{fig13}, we reproduce the result 
of Ref.~\cite{Bedaque:1998km}, displaying the approximate RG limit cycle with quasi-periodic singularities 
associated with the successive formation of new Efimov states as $\Lambda_{\rm reg}\to \infty$. The data points 
correspond to our numerical evaluations, while the solid curve is the fit to these data using the analytical 
formula of Bedaque {\it et al.}, Eq.~\eqref{EQ23}. To this end one may extract the three-body parameters, such as
$\Lambda_*$ and $s_0$, using the momentum scaling relations, 
\begin{equation*}
  \Lambda_n=\left(e^{\pi/s_0}\right)^n \Lambda_0 \quad \text{as}\,\, n\to \infty  \,,
\end{equation*}
and 
\begin{equation}
  \Lambda_0= \exp\left[\frac{\arctan (1/s_0)}{s_0}\right] \Lambda_* \,,     
\end{equation}
where $\Lambda_{\rm reg}= \Lambda_n$ represents the $n^{\rm th}$ zero of the three-body coupling $H(\Lambda_n)$. 
The typical non-asymptotic values of $s_0$ expected in this case depend on $\Lambda_n$, and hence differ from the 
cutoff independent asymptotic value, $s_0 \to 1.00624\cdots$, which restores the exact scaling symmetry of the STM 
equation as $\Lambda_n\to \infty$.  

%%%%%%%%%%%%%%%%%%%%%%%%%%%%%%%%%%%%%%%%%%%%%%%%%%%%%%%%%%%%%%%%%%%%%%%%%%%%%%%%%%%%%%%%%%%%%%%%%%%%%%%%%%%%%%%%%%%%
\section{\,\,\,Integral equation for $\Xi^- nn$ ($I=3/2, J^P={1/2}^+$) system}
%%%%%%%%%%%%%%%%%%%%%%%%%%%%%%%%%%%%%%%%%%%%%%%%%%%%%%%%%%%%%%%%%%%%%%%%%%%%%%%%%%%%%%%%%%%%%%%%%%%%%%%%%%%%%%%%%%%%
The Faddeev-like three-body coupled integral equation for the $n+(\Xi^- n)_t\to n+(\Xi^- n)_t$ elastic scattering 
amplitude $t_A$ (cf. Fig.~\ref{fig-1}) (excluding the $(\Xi^-n)_s$ singlet subsystem channel) can be easily obtained 
{\it via} Feynman rules from the non-relativistic effective Lagrangian, and is given as 
%%%%%%%%%%%%%%%%%%%%%%%%%%%%%%%%%%%%%%%%%%%%%%%%%%%%%%%%%%%%%%%%%%%%%%%%%%%%%%
\begin{eqnarray}
 t_A \left({\bf p,k};E\right) &=& 
 (-y_1^2)\left[{\cal C}^{(11)}_2 S_{\Xi}\left(E-\frac{{\bf p}^2}{2M_n}-\frac{{\bf k}^2}{2M_n},{\bf p+k}\right) 
 + {\cal C}^{(11)}_3\frac{M_{\Xi}}{2}\frac{g_3(\Lambda_{\rm reg})}{\Lambda^2_{\rm reg}} \right]
 \nonumber\\
 && -\, {\cal C}^{(11)}_2 (-y_1^2)\int^{\Lambda_{\rm reg}} \frac{{\rm d}^3{\bf q}}{(2\pi)^3}\,
 t_A\left({\bf q,k};E\right)\, S_{\Xi}\left(E-\frac{{\bf p}^2}{2M_n}-\frac{{\bf q}^2}{2M_n},{\bf p+q}\right)
 \nonumber\\
 && \hspace{4cm} \times\,\mathscr{D}_1\left(E-\frac{{\bf q}^2}{2M_n},{\bf q}\right)
 \nonumber \\
 && -\, {\cal C}^{(11)}_3(-y^2_1)\frac{M_{\Xi}}{2} \int^{\Lambda_{\rm reg}} \frac{{\rm d}^3{\bf q}}{(2\pi)^3}
 t_A\left({\bf q,k};E\right) \mathscr{D}_1\left(E-\frac{{\bf q}^2}{2M_n},{\bf q}\right)
 \frac{g_3(\Lambda_{\rm reg})}{\Lambda^2_{\rm reg}}
 \nonumber\\
 && -\, {\cal C}^{(10)}_2 (-y_0y_1)\int^{\Lambda_{\rm reg}} \frac{{\rm d}^3{\bf q}}{(2\pi)^3}\,
 t_B\left({\bf q,k};E\right) S_n\left(E-\frac{{\bf p}^2}{2M_n}-\frac{{\bf q}^2}{2M_{\Xi}},{\bf p+q}\right)
 \nonumber\\
 && \hspace{4cm} \times\,\mathscr{D}_0\left(E-\frac{{\bf q}^2}{2M_{\Xi}},{\bf q}\right)
 \nonumber\\
 && +\, {\cal C}^{(10)}_3 (-y_0y_1)\sqrt{\frac{3}{2}}M_n \int^{\Lambda_{\rm reg}}\,\frac{{\rm d}^3{\bf q}}{(2\pi)^3}\, 
 t_B\left({\bf q,k};E\right) \frac{g_3(\Lambda_{\rm reg})}{\Lambda^2_{\rm reg}}
 \nonumber\\
 && \hspace{4cm} \times\,\mathscr{D}_0\left(E-\frac{{\bf q}^2}{2M_{\Xi}},{\bf q}\right)\,,
 \label{EQ29}
\end{eqnarray}
and 
%%%%%%%%%%%%%%%%%%%%%%%%%%%%%%%%%%%%%%%%%%%%%%%%%%%%%%%%%%%%%%%%%%%%%%%%%%%%%%
\begin{eqnarray}
 t_B\left({\bf p,k};E\right) &=& 
 (-y_0y_1)\left[{\cal C}^{(10)}_2  S_n\left(E-\frac{{\bf k}^2}{2M_n}-\frac{{\bf p}^2}{2M_{\Xi}},{\bf p+k}\right) 
 - {\cal C}^{(10)}_3 \sqrt{\frac{3}{2}}M_n\frac{g_3(\Lambda_{\rm reg})}{\Lambda^2_{\rm reg}} \right]
 \nonumber\\
 &&-\, {\cal C}^{(10)}_2(-y_0y_1) \int^{\Lambda_{\rm reg}}\frac{{\rm d}^3{\bf q}}{(2\pi)^3}
 t_A\left({\bf q,k};E\right) S_n\left(E-\frac{{\bf p}^2}{2M_{\Xi}}-\frac{{\bf q}^2}{2M_n},{\bf p+q}\right)
 \nonumber\\
 && \hspace{4cm} \times\, \mathscr{D}_1\left(E-\frac{{\bf q}^2}{2M_n},{\bf q}\right)
 \nonumber\\
 &&+\, {\cal C}^{(10)}_3 (-y_0y_1)\sqrt{\frac{3}{2}}M_n \int^{\Lambda_{\rm reg}}\frac{{\rm d}^3{\bf q}}{(2\pi)^3}\,
 t_A\left({\bf q,k};E\right)\frac{g_3(\Lambda_{\rm reg})}{\Lambda^2_{\rm reg}}
 \nonumber\\
 && \hspace{4cm} \times\,\mathscr{D}_1\left(E-\frac{{\bf q}^2}{2M_n},{\bf q}\right)\,.
 \label{EQ30}
\end{eqnarray}
%%%%%%%%%%%%%%%%%%%%%%%%%%%%%%%%%%%%%%%%%%%%%%%%%%%%%%%%%%%%%%%%%%%%%%%%%%%%%%
Here, ${\cal C}^{(11)}_2= 1/2$, ${\cal C}^{(10)}_2=-\sqrt{3/2}$ are spin-isospin re-coupling coefficients for diagrams 
with two-body interaction only, and ${\cal C}^{(11)}_3={\cal C}^{(10)}_3=1$ are those with the three-body contact 
interaction. Upon renormalization using the wavefunction renormalization constant $\mathcal Z_{\Xi n}$ (cf. 
Eq.~\eqref{EQWFZ}), and projecting on to the S-wave, the above amplitudes lead to Eqn.~\eqref{eq:11} and \eqref{eq:12}.

\end{appendix}

%%%%%%%%%% REFERENCES %%%%%%%%%%%%%%%%%%%%%%%%%

\end{document}